\documentclass[review]{elsarticle}

\usepackage{hyperref}
\usepackage{graphicx}
\usepackage[utf8]{inputenc}
\usepackage{amssymb}
\usepackage{makecell}
\usepackage[figuresleft]{rotating}
\usepackage{multirow}
\usepackage[table,xcdraw]{xcolor}
\usepackage{pifont}
\usepackage{adjustbox}
\usepackage{amsmath}
\usepackage[linesnumbered,ruled,vlined]{algorithm2e}

\usepackage{chngpage}

\usepackage{tabularx}
\usepackage[table]{xcolor}
\usepackage[draft]{changes} 
\usepackage{xcolor}
\usepackage{array}
\usepackage{ltablex,array}
\newcommand{\PreserveBackslash}[1]{\let\temp=\\#1\let\\=\temp}
\newcolumntype{C}[1]{>{\PreserveBackslash\centering}p{#1}}
\journal{Journal of Systems and Software}

\begin{document}
	
	\begin{frontmatter}
		
		\title{GloBug: Using Global Data in Fault Localization}
		
		\author{Nima Miryeganeh\corref{cor2}}

		\ead{seyednima.miryeganeh@ucalgary.ca}
		
		\author{Sepehr Hashtroudi\corref{cor1}}		
		\ead{sepehr.pourabolfathh@ucalgary.ca}

		\author{Hadi Hemmati\corref{cor2}}
		\ead{hadihemmati@ucalgary.ca}
		\address{Department of Electrical and Computer Engineering  \\
			Schulich School of Engineering, University of Calgary, Canada}
		
		



		\cortext[cor1]{Corresponding author}
		
		\begin{abstract}
			Fault Localization  (FL) is  an  important  first step  in  software  debugging and is mostly manual in the current practice. Many methods have been proposed over years to automate the FL process, including information retrieval (IR)-based techniques. These methods localize the fault based on the similarity of the reported bug report and the source code.  Newer variations of IR-based FL (IRFL) techniques also look into the history of bug reports and leverage them during the localization. However, all  existing IRFL techniques limit themselves to the current project's data (local data). In this study, we introduce $Globug$, which is an IRFL framework consisting of methods that use models pre-trained on the global data (extracted from open-source benchmark projects). In $Globug$, we investigate two heuristics: a) the effect of global data on a state-of-the-art IR-FL technique, namely $BugLocator$, and b) the application of a Word Embedding technique (Doc2Vec) together with global data. 
			Our large scale experiment on 51 software projects shows that using global data improves $BugLocator$ on average 6.6\% and 4.8\% in terms of MRR (Mean Reciprocal Rank) and MAP (Mean Average Precision), with over 14\% in a majority (64\% and 54\% in terms of MRR and MAP, respectively) of the cases. This amount of improvement is significant compared to the improvement rates that five other state-of-the-art IRFL tools provide over $BugLocator$. In addition, training the models globally is a one-time offline task with no overhead on $BugLocator$'s run-time fault localization. Our study, however, shows that a Word Embedding-based global solution did not further improve the results. 
		\end{abstract}
		
		\begin{keyword}
			Automated Fault Localization \sep
			Information Retrieval\sep
			Word Embedding\sep
			TF.IDF\sep
			Doc2Vec\sep
			Global training
		\end{keyword}
		
	\end{frontmatter}

	\section{Introduction}
	\label{Introduction_section}
	
	Software debugging is one of the most essential but costly activities in software development and maintenance \cite{debug1,debug2,debug3}, which involves locating a bug, understanding the issue, and fixing it.
	
	Fault Localization (FL) refers to the process of locating the program elements (functions, classes, files, etc.), which are associated with a fault \cite{FL1,LDA2,FL3,FL4}. Nowadays in practice, most of the software debugging processes is done manually by the testers/developers who are responsible for finding/fixing the reported bug. This process, however, is very time-consuming, especially for larger software systems. 
	
	There has been a great number of studies in the past two decades that have focused on automating the FL process. 
	The Automated FL techniques use heuristics to determine which program elements are most suspicious and most likely to be associated with a fault. Hence, a programmer saves time during debugging by focusing attention on the most suspicious locations \cite{survey4}.
	
	There are several automated FL techniques in the literature such as Spectrum-based Fault Localization (SFL), Information Retrieval-based Fault Localization (IRFL), Mutation-based, and Model-based Diagnosis.  
	
	IRFL techniques, which are the context of this study, mostly use the static information provided in bug reports. For example the title, the full description, etc. They treat a bug report as a query and rank the source code files by their relevance to the query. The developers then
	examine the returned files and fix the bug. IRFL approaches typically have lower computational cost and require minimal information (e.g., requiring only source code and bug reports to operate) compared to alternatives \cite{IRFLLowCost}. However, in general, no one method outperforms all others in all systems \cite{Bench4Bl, survey4}. 


	A typical IRFL technique uses a relevancy function to rank the relevant program elements. 
	The relevancy functions calculate the textual similarity of the new bug report to other historical bug reports or source code elements, and use this information to score most similar program elements to the current bug report. Therefore in the core of any IRFL technique there should be a textual similarity function that accepts two documents and returns a similarity value. 
	
	In IR, there has been a wide range of textual similarity functions proposed over the years to rank a corpus of documents based on their relevance to a query. These techniques usually calculate the frequency (in one way or another) of the common terms that appear in the query and the document. Most technique, first represent the documents as a vector of numbers. Then they apply an standard vector similarity/distance measure (e.g. cosine) on them. In most IRFL literature the vectorization algorithm is TF.IDF \cite{tfidf}. TF.IDF method will be explained in details in Section\ref{BugLocator_Section}.

	\color{black}
	
	To improve the accuracy of the existing IRFL methods, in this paper, we introduce $Globug$, an IRFL technique that implements two novel heuristics: 
	\begin{enumerate}
		\item Leveraging a global dataset (which is an existing benchmark dataset of bug reports in several open source projects) as part of the TF.IDF model training within a given project, to better calculate textual similarities.
		\begin{itemize}
			\item \textbf{Motivation:} To use the freely available knowledge outside of the current project to better calculate the textual similarities, when the current project's training set (historical bug reports) is not rich enough. Note that in general, building vocabulary mostly benefits from larger dataset, since tokens that are not relevant to the current project will be automatically ignored in the process of similarity calculation.   
		\end{itemize}
		\color{black}
		
		\item Using a word embedding-based language model (Doc2Vec\cite{Doc2Vec}) to make a better language model out of the global data. 
		\begin{itemize}
			\item \textbf{Motivation:}  The fact that recent neural embedding techniques such as Doc2Vec improve the semantic capturing ability of classic vectorization approaches like TF.IDF, in many domains,  motivates us to replace TF.IDF with Doc2Vec. Also note that the large dataset (global data) is the key enabler for applying embedding models, which could not be done if we would only look at one project's history. 
		\end{itemize}
		\color{black}
		
	\end{enumerate}
	Our main comparison baseline is $BugLocator$ \cite{bugLocator}, which has served as baseline TF.IDF-based FL technique in many IRFL literature. However, we also compare our improvements over $BugLocator$ with improvements that other more recent tools (five recent IRFL tools) provide over $BugLocator$, and discuss our advantages over them. 
	
	To evaluate the above heuristics, we have designed and reported an empirical study on 51 software projects from an open source benchmark dataset for IRFL techniques \cite{Bench4Bl}, as follows:
	\begin{itemize}
		\item Heuristic1: How does a global corpus affect $BugLocator$? 
		Our results suggest that using global corpus for TF.IDF improves $BugLocator$ with average rates of 6.6\% and 4.8\%, in terms of MRR and MAP. The MRR and MAP improvements were more than 14\% in 64\% and 54\% of the cases, respectively. Looking at the other more recent IRFL tools that improve $BugLocator$ shows that our improvement is relatively significant.
		
		\item Heuristic2: Can Word Embedding improve TF.IDF? 
		Our key findings for this heuristic is that an advance Word Embedding technique may not always result in improvement over TF.IDF, and its application in IRFL must be accompanied with caution.
	\end{itemize}
	\color{black}
	In terms of overhead cost of using a global corpus, we also argue that our approach is quite light-weight, given that the extra calculations (either for TF.IDF or the Embedding model) over $BugLocator$ is done once, offline, and will be reused for all new bugs of a project. 
	Note that the Bench4Bl benchmark already exists and can be used for model training once and reused for any new project outside of these 51 projects as well (that is no extra effort is needed if $Globug$ is applied on a new project). Of course, enriching the benchmark in the future with more projects may improve the results, but it is not a requirement for $Globug$ and the reported performance of $Globug$ in this paper is .
	
	In summary, the contributions of this paper are:
	
	\begin{itemize}
		\item Proposing a new light-weight idea of leveraging freely available benchmark datasets (global corpus) such as Bench4BL \cite{Bench4Bl} during an offline training period in TF.IDF-based IRFL techniques. 
		\item Proposing the new idea of replacing TF.IDF with a Word-Embedding model (such as Doc2Vec) in IRFL techniques and pre-train the embedding on the global corpus.
		\item Conducting a large scale empirical study on the effectiveness of TF.IDF vs. Doc2Vec Embedding-based FL solutions, trained and evaluated on the global corpus of 51 open source projects (50 projects for training + 1 for testing, at each run), from Bench4BL \cite{Bench4Bl}, including a detail investigation of the embedding results. 
	\end{itemize}
	
	We also provide the replication package as well as all raw results available online. \footnote{\href{https://github.com/miryeganeh/GloBug}{https://github.com/miryeganeh/GloBug}}

	The rest of this paper is organized as follows: in Section \ref{background_section} IRFL techniques in general and $BugLocator$ as our comparison baseline in particular will be explained. We also briefly explain the basics of the embedding technique used in this paper. 
	In Section\ref{Methodology_section} 
	we depict the big picture of our proposed framework and our two heuristics and explain the details of their inner processes. 
	In Section \ref{Empirical Evaluation_section}, we explain the design and results of our experiments. In Section \ref{discussion_section}, we discuss the results and analyze possible threats to their validity by taking a deeper look at some specific examples.
	Next, in Section \ref{RW_section}, we take a look at the highlights of related IRFL studies and position this paper among them.
	In Section \ref{conclusion_section}, we analyze the possible future paths of using global data and Word Embedding in IRFL and the potential extensions of this study. 
	Finally, in Section \ref{conclusion_section}, we conclude our study by summarizing the experiments and findings.

	\section{Background}
	\label{background_section}
	
	
	
	There has been a wide range of research studies conducted over the past two decades in FL \cite{WongSurvey}. There are at least two leading directions of research among these studies, namely spectrum-based fault localization \cite{SFL1,SFL2,SFL3,SFL4,SFL5,zhang2019empirical,jiang2019combining,wen2019historical,he2020enhancing} and IR-based fault localization
	\cite{LDA1,bugLocator,Locus,HyLoc,Amalgam,BLUiR,BLIA,BRTracer, rahman2018improving, mills2020relationship, mills2018bug}. In the former, the likely locations of faults are identified by computing some ranking metrics, generally based on similarity coefficients and statistical techniques, on succeeding and failing test execution traces. The latter approaches, on the other hand, only leverage static properties of the software (e.g. source code files and bug reports and no test execution) to identify suspicious files, using IR techniques. Given that the context of this paper is about IRFL techniques and for the sake of brevity, we skip the background on SFL.

	\subsection{IR-Based Fault Localization (IRFL)}
	\label{IR-Based Fault Localization (IRFL)_section}
	
	In IRFL methods bug reports play an essential role in localization, as they usually contain a detailed description of the failure, and occasionally give valuable hints on the location of the fault in the software. 
	In IRFL, the goal is to bridge from a bug report to buggy elements of the program. Therefore, the problem can be transformed into an IR problem \cite{IR} in which documents (program elements) are ranked based on their relevancy to a query (bug report). 
	
	\begin{figure}[t]
		\includegraphics[width=\linewidth]{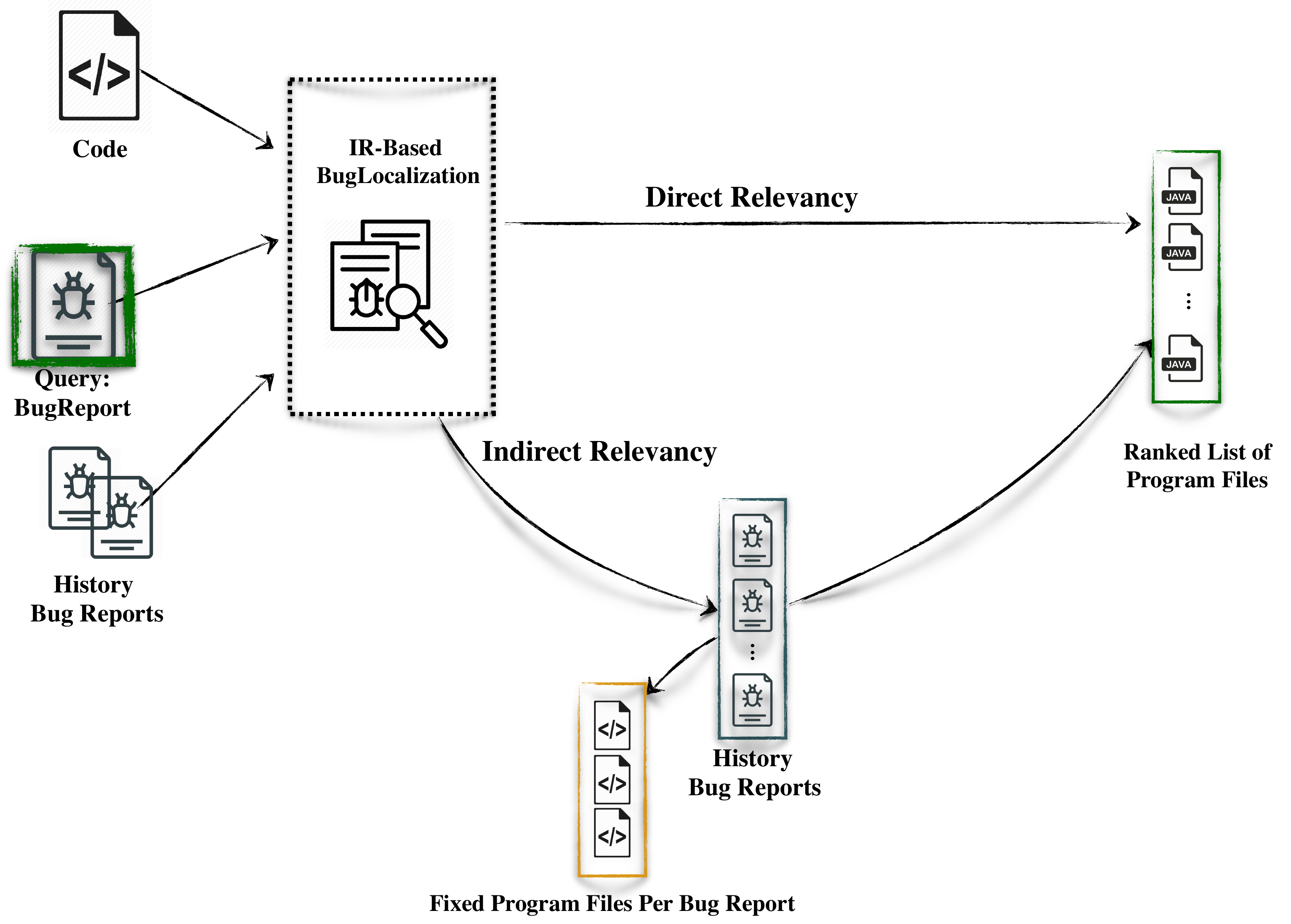}
		\caption{High-level process of direct and indirect relevancy functions in IRFL.}
		\label{Direct Relevancy VS. Indirect Relevancy}
	\end{figure}
	
	As depicted in Figure \ref{Direct Relevancy VS. Indirect Relevancy}, the relevance of program elements to bug reports can be calculated directly or indirectly using a relevancy function: 
	
	\begin{itemize}
		\item {Direct Relevancy Function}: A direct relevancy function calculates the relevancy score of program elements and bug reports (query) using different heuristics, which are mostly based on the direct textual similarity of bug reports and program elements. 
		
		\item {Indirect Relevancy Function}: An indirect relevancy function calculates the relevancy score, first by calculating the similarity of the bug report (query) to historical bug reports in the software project, and then knowing the buggy elements that have been fixed for each historical bug report, calculates an indirect relevancy score for each program element. 
	\end{itemize}
	
	$BugLocator$ \cite{bugLocator} is one of the most well-known IRFL techniques proposed by Jian Zhou et. al., in 2012. $BugLocator$ incorporates both the direct and indirect relevance of bug reports to source code files to localize buggy source files in the system. $BugLocator$ can be considered as a turning point in IRFL studies as it had a significantly better performance compared to its older competitors, and it is now often used by other IRFL techniques as a state-of-the-art baseline method. Therefore, in the following section, we first take a deeper look at the internal structure of the $BugLocator$. 
	
	\subsection{BugLocator - Our Baseline Method}
	\label{BugLocator_Section}
	
	
	\begin{figure}[!t]
		\centering
		\includegraphics[width=0.9\linewidth]{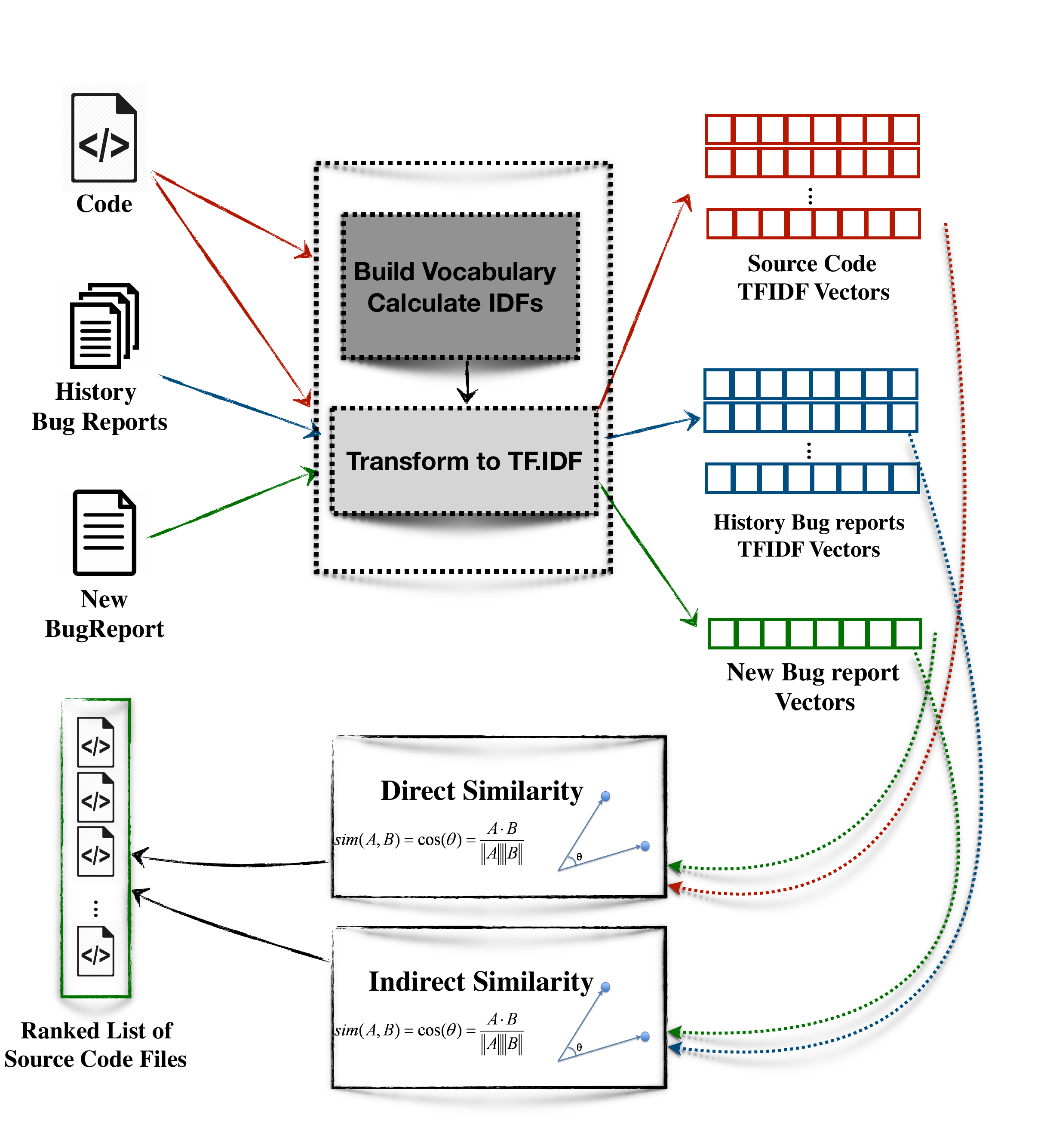}
		\caption{Illustration of internal process of $BugLocator$}
		\label{BugLocator}
	\end{figure}
	
	Figure \ref{BugLocator} summarizes the internal process of $BugLocator$. In $BugLocator$, first, a shared vocabulary is formed using all the distinct words that appear in all source code files and bug reports. Then, an IDF (inverse document frequency) weight is calculated for each word based on the frequency of that word in all documents. 
	Next, a vector of TF.IDF (term frequency . inverse document frequency) weights is calculated for each bug report and source code file based on the product of the frequency of each vocabulary word (TF) in that document and the IDF of the vocabulary word that denotes the importance of the word in the context.
	Therefore, a common coding term such as $println$ (print line in Java) that appears in many source code files (high document frequency) will get a lower IDF (lower importance), which in turn decreases the impact of its TF.IDF weight while calculating the similarity scores.
	
	$BugLocator$ uses an altered version of the TF.IDF formula (Equation \ref{rVSM}) namely rVSM (revised Vector Space Model) to calculate the similarity score, which takes the length of the source code file into account too. The authors claim that it works best in their context compared to other variations.
	
	The rVSM function calculates the relevancy of a source Code file to a bug report among all other source code files using the cosine similarity of their TF.IDF vectors.
	
		\begin{equation}
		\begin{aligned}
		&rVSM(BR,SC) = \frac{1}{1+e^{-N(\#terms)}}\\
		&\times \frac{1}{\sqrt{\sum_{t\in BR}((\log{f_{t_{BR}}}+1)\times\log{(\frac{\#docs}{n_{t}})})^{2}}}\\
		&\times \frac{1}{\sqrt{\sum_{t\in SC}((\log{f_{t_{SC}}}+1)\times\log{(\frac{\#docs}{n_{t}})})^{2}}}\\
		&\times \sum_{t\in{BR} \cap{SC}}(\log{(f_{t_{BR}}}+1)\times(\log{f_{t_{SC}}}+1)\\
		&\times\log{(\frac{\#docs}{n_{t}})^2})
		\end{aligned}
		\label{rVSM}
		\end{equation}
		where $N(\#terms)$ refers to the normalized value of the number of terms in the document, $f_{t_{BR}}$ refers to the number of occurrences of a term t in the bug report, $f_{t_{SC}}$ refers to the number of occurrences of a term t in the source code, $n_{t}$ refers to the number of documents that contain the term t, $\#docs$ represents the total number of documents in the corpus.


	The lexical gap between bug reports and source code, sometimes makes the direct comparison unreliable. Therefore, in the next step, $BugLocator$ calculates an ``indirect relevancy'' score for each source code file.
	
	
	The indirect relevancy is designed to help the direct relevancy in cases where the bug report does not contain many source code-related terms. In such cases, the indirect relevancy help find the relevant program elements using the similarity of the language used in the new bug report and the history bug reports. $BugLocator$ linearly combines these two relevancy functions as its final relevancy function. 

	\subsection{Neural Word Embedding}
	In $BugLocator$ both direct and indirect relevancy functions represent a document (bug report or a source code file) as a vector of TF.IDF values assigned to each word, within the document. Word embedding is an alternative way to vectorize the documents. There are many embedding techniques in the literature, but Word2Vec, introduced by Mikolov et. al. in 2013 at Google \cite{Word2Vec} attracted great attention of researchers and companies. Word2Vec trains a shallow (2-layer) neural network to learn which words are most likely appear together in a document (i.e., semantic relationship) and represents them as close vectors in an N-dimensional space. Doc2Vec \cite{Doc2Vec} used in this paper is an extension of Word2Vec where the vectors represent the entire document rather than individual words, which is more efficient given that an FL relevancy function only cares about distances between documents (bug and code) and not the individual words. 
	More formally, given a sequence of training words
		$w_1 ,w_2 ,w_3 ,...,w_T$ , the objective of the word vector model is to maximize the average log probability. 
		\begin{equation}
		\begin{aligned}
		\frac{1}{T}\sum_{t=k}^{T-k}\log p(w_t|w_{t-k}, ...,w_{t+k})
		\end{aligned}
		\end{equation}
		The prediction task is typically done via a multiclass classifier, such as softmax. There, we have
		\begin{equation}
		\begin{aligned}
		p(w_t|w_{t-k}, ...,w_{t+k}) = \frac{e^{y_{w_t}}}{\sum_{i} e^{y_i}}
		\end{aligned}
		\end{equation}
		Each of $y_i$ is un-normalized log-probability for each output word $i$, computed as
		\begin{equation}
		\begin{aligned}
		y = b + Uh(w_{t-k}, ..., w_{t+k};W)
		\end{aligned}
		\end{equation}
		where $U,b$ are the softmax parameters. $h$ is constructed by a concatenation or average of word vectors extracted from $W$.

	\section{$Globug$ - Our Proposed IRFL Method}
	\label{Methodology_section}
	
	
	
	$Globug$'s main proposal is using a global corpus of benchmark projects during the training phase of language model building. We explore this idea in two variations:
	
	\subsection{$Globug$ -  Variation 1: TF.IDF and Global Data}
	
	In the first variation, the aim is reusing as much as possible from $BugLocator$ and only replace the local training data (current project) to the global corpus. That means that we still use TF.IDF as the vectorizer, but calculate IDFs over the global dataset. The motivation behind this idea is that if IDF weights are calculated over a much bigger corpus of documents, the importance of terms are more precisely captured. 
	
	
	
	\begin{figure}[!t]
		\centering
		\includegraphics[width=\linewidth]{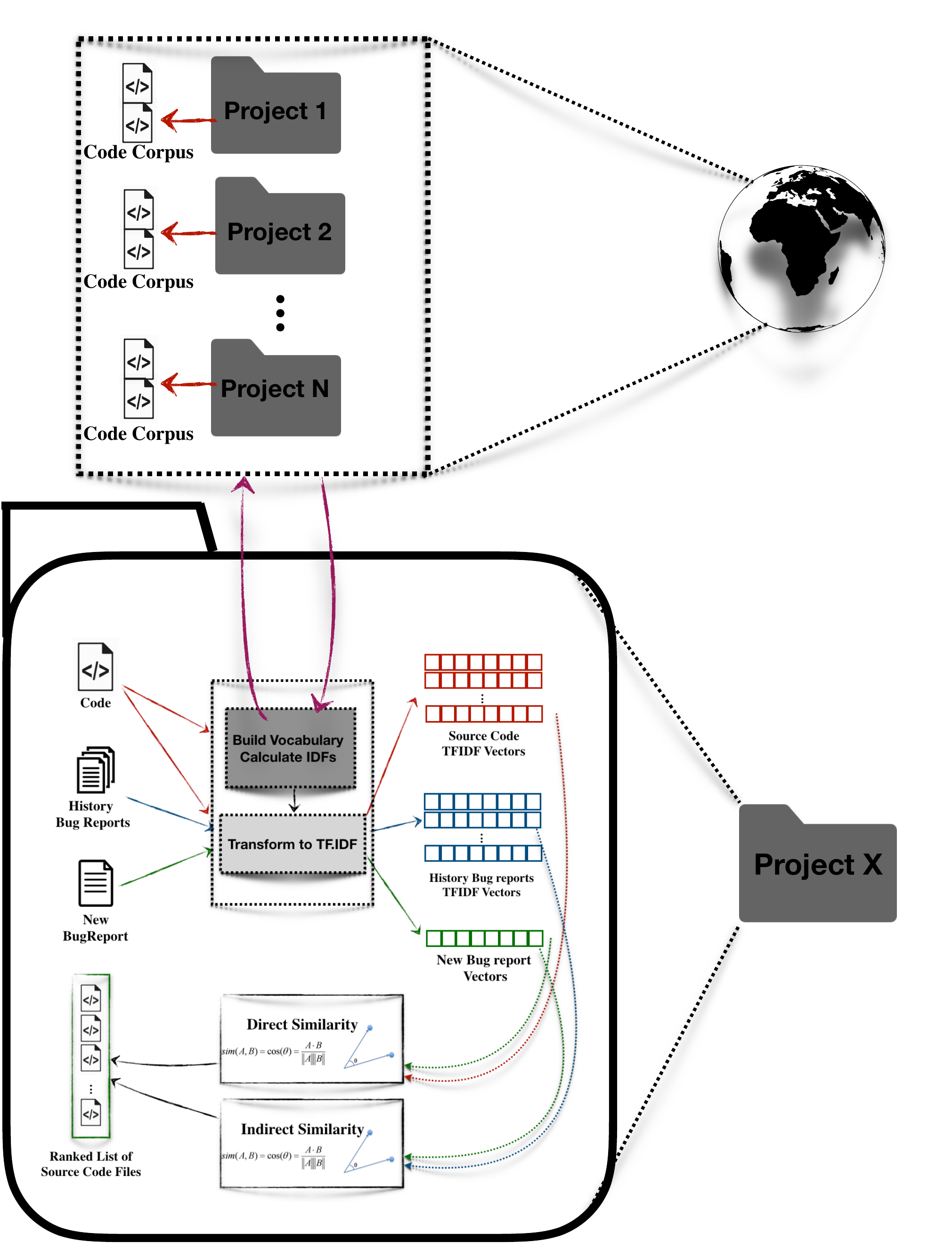}
		\caption{Illustration of $Globug$ Process -  Variation 1: TF.IDF and Global Data}
		\label{Global BugLocator}
	\end{figure}
	

	

	In $Globug$ (Variation 1), the global corpus is used for both direct and indirect relevancy functions, as follows:
	
	\begin{itemize}
		\item \textbf{Direct Relevancy}: As demonstrated in Figure \ref{Global BugLocator}, first the vocabulary with all unique terms of all $N$ projects (global corpus) source code files is created.  Then $IDF$ of all unique terms over the global corpus projects are calculated.  Note that since the dictionary is built only using source code files of all projects, any terms in bug reports that are not seen in the dictionary are ignored. Finally, the direct relevancy score of a source code file to a bug report is calculated based on the cosine similarity of their globally calculated $TF.IDF$ vectors.
		
		\item \textbf{Indirect Relevancy}: Similar to $BugLocator$, an indirect relevance score is calculated, first, by relating the history bug reports to the new bug report and then bridging from history bug reports to source code files. However, the difference is that in $Globug$ (Variation 1), the dictionary is built using common source code terms of all projects and their $IDF$ is calculated globally. 
	\end{itemize}
	
		As you see in algorithm \ref{globug_var_1}, first a dictionary is created using unique words of all projects source codes. Then the $TF.IDF$ vectors are calculated for All Source Code files(ASC), History of Bug Reports(HBR), and the New Bug Report(NBR). Afterwards, for the direct relevancy, the Cosine Similarity($COSIM_{DR}$) of the $TF.IDF_{NBR}$ and $TF.IDF_{ASC}$ is calculated, And for the indirect relevancy, the Cosine Similarity($COSIM_{IDR}$) of the $TF.IDF_{NBR}$ and $TF.IDF_{HBR}$ is calculated. At the end, for generating the Ranked List of Source Codes($RLSC$) we combine direct and indirect relevancy scores using a weighted average stated in equation \ref{weighted_average}. The value of $W_1$ and $W_2$ in our experiments are 0.8 and 0.2, respectively, which is the same as $BugLocator$.
		\begin{equation}
		SimilarityScore(file) = COSIM_{DR}(file)*W_1 + COSIM_{IDR}(file)*W_2
		\label{weighted_average}
		\end{equation}
		\begin{algorithm}[H]
			\KwData{ASC: All projects Source Codes}
			\KwData{HBR: History of Bug Reports}
			\KwData{NBR: New Bug Report}
			\KwResult{RLSC: Ranked List of Source Code files}
			\For{NextNewWord W in ASC} {
				$Dict_{SC} \leftarrow W$\tcp*[1]{Make dictionary}
				$IDF.Model \leftarrow IDF(W,ASC)$\tcp*[1]{Calculate IDF of docs}
			}
			\For{AllDocs D in (ACS,HBR,NBR)}{
				$TF.IDF_{ACS} \leftarrow TF.IDF.Model(D)$\;
				$TF.IDF_{HBR} \leftarrow TF.IDF.Model(D)$\;
				$TF.IDF_{NBR} \leftarrow TF.IDF.Model(D)$\;
			}
			$COSIM_{DR} \leftarrow  CosineSimilarity(TF.IDF_{NBR}, TF.IDF_{ACS})$\;
			$COSIM_{IDR} \leftarrow  CosineSimilarity(TF.IDF_{NBR}, TF.IDF_{HBR})$\;
			$COSIM = COSIM_{DR}*W_1 + COSIM_{IDR}*W_2$\;
			$RLSC \leftarrow ArgMax(COSIM)$\;
			\caption{$Globug$ algorithm (Variation 1)}
			\label{globug_var_1}
		\end{algorithm}

	
	\subsection{$Globug$ -  Variation 2: Word Embedding and Global Data}
\begin{figure}[!t]
		\includegraphics[width=\linewidth]{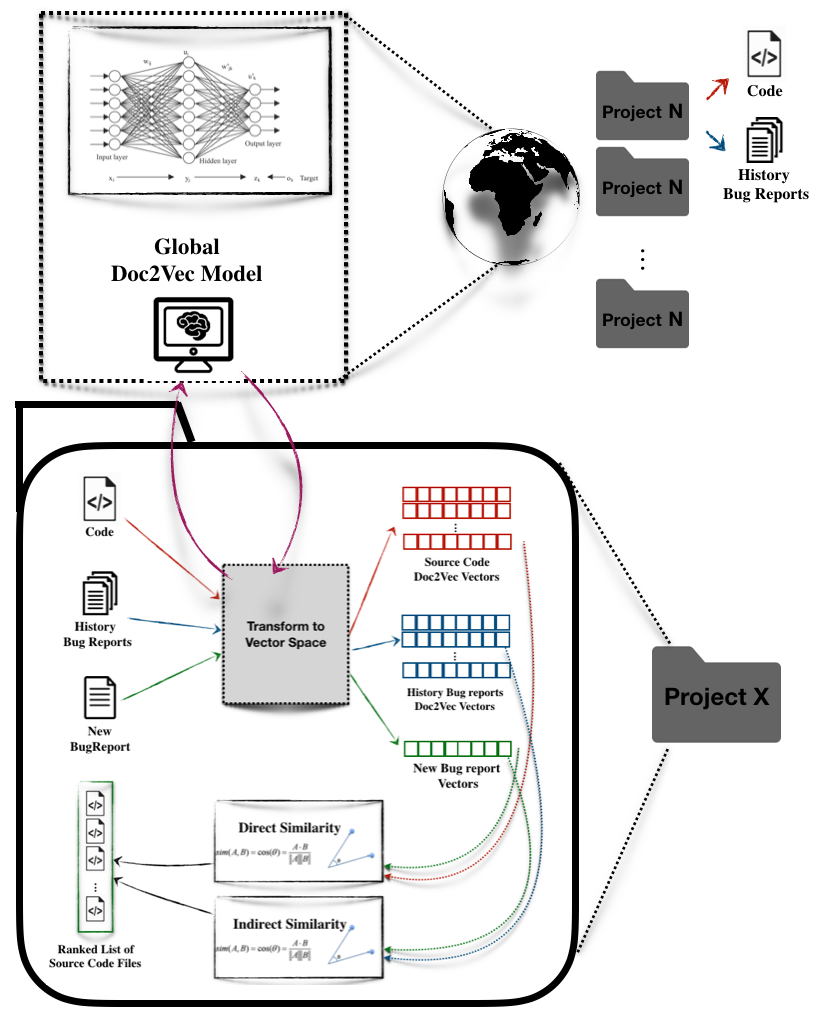}
		\caption{Illustration of $Globug$ Process -  Variation 2: Word Embedding and Global Data}
		\label{GlobalDoc2Vec}
	\end{figure}

	In the second variation of $Globug$, the main goal is to replace TF.IDF vectorization with a Doc2vec embedding and we still use the global dataset as the training set. The motivation and expectation from Doc2Vec is to better capture the semantic relationship between the documents. Figure \ref{GlobalDoc2Vec} illustrates the process in details. Note that the two variations look exactly the same in terms of FL process except that one vectorizes the documents using globally calculated TF.IDFs and the other using globally calculated Doc2Vec. 
	
		As you see in algorithm \ref{globug_var_2}, first a dictionary is created using unique words of all projects source codes. Then the $Doc2Vec$ model is trained on the dictionary. Then $Doc2Vec$ and $TF.IDF$ vectors are calculated for All Source Code files(ASC), History of Bug Reports(HBR), and the New Bug Report(NBR). Afterwards, for the direct relevancy, the Cosine Similarity($COSIM_{DR}$) of the $TF.IDF_{NBR}$ and $TF.IDF_{ASC}$ is calculated, And for the indirect relevancy, the Cosine Similarity($COSIM_{IDR}$) of the $Doc2Vec_{NBR}$ and $Doc2Vec_{HBR}$ is calculated.\vspace{.5cm}\\
		\vspace{.5cm}
		\begin{algorithm}[H]
			\KwData{ASC: All projects Source Codes}
			\KwData{HBR: History of Bug Reports}
			\KwData{NBR: New Bug Report}
			\KwResult{RLSC: Ranked List of Source Code files}
			\For{NextNewWord W in ASC} {
				$Dict_{SC} \leftarrow W$\tcp*[1]{Make dictionary}
				$IDF.Model \leftarrow IDF(W,ASC)$\tcp*[1]{Calculate IDF of docs}
				$Doc2Vec.Model \leftarrow Doc2Vec(W,ASC)$\tcp*[1]{Train Doc2Vec model}
			}
			\For{AllDocs D in (ACS,HBR,NBR)}{
				$Doc2Vec_{HBR} \leftarrow Doc2Vec.Model(D)$\;
				$TF.IDF_{ACS} \leftarrow TF.IDF.Model(D)$\;
				$Doc2Vec_{NBR} \leftarrow Doc2Vec.Model(D)$\;
				$TF.IDF_{NBR} \leftarrow TF.IDF.Model(D)$\;
			}
			$COSIM_{DR} \leftarrow  CosineSimilarity(TF.IDF_{NBR}, TF.IDF_{ACS})$\;
			$COSIM_{IDR} \leftarrow  CosineSimilarity(Doc2Vec_{NBR}, Doc2Vec_{HBR})$\;
			$COSIM = COSIM_{DR}*W_1 + COSIM_{IDR}*W_2$\;
			$RLSC \leftarrow ArgMax(COSIM)$\;
			\caption{$Globug$ algorithm (Variation 2)}
			\label{globug_var_2}
		\end{algorithm}
		At the end, for generating the Ranked List of Source Codes($RLSC$) we combine direct and indirect relevancy scores using a weighted average stated in equation \ref{weighted_average}. The value of $W_1$ and $W_2$ in our experiments are 0.8 and 0.2, respectively (the same as $Globug$ -  Variation 1).
	\\
	In terms of implementation, we reuse TF.IDF implementation from $BugLocator$ and also reuse a commonly used implementation of Doc2Vec\cite{Doc2Vec}. 
	
	Note that in our study, the existence of Bench4BL \cite{Bench4Bl} made the experimentation part easier, however, Bench4BL is just one example of global corpus for FL. In fact, any extension of Bench4BL or any other (larger or more diverse) benchmark datasets that are created in the future can be used in $Globug$ and potentially improve its results. However, in practice one does not necessarily need to come up with new ``global corpuses'' or improve Bench4Bl to be able to apply $Globug$ in their project.

	\section{Empirical Study}
	\label{Empirical Evaluation_section}
	In this section, we explain our experiments to evaluate the effectiveness of $Globug$.

	\subsection{Objectives and Research Questions}
	\label{Research Questions_section}
	
	The objective of this study is to investigate the effect of our two heuristics a global corpus and Word Embedding on IRFL, to answer the following five research questions (RQs): 
	
	\begin{itemize}
		\item Heuristic1: How does a global corpus affect $BugLocator$? 
		\begin{itemize}
			\item RQ1. Can a global corpus improve effectiveness of the direct relevancy function in $BugLocator$? 
			
			The RQ1 goal is to see the effect of the global corpus only on the direct-relevancy function, when there is no indirect-relevancy included.
			
			\item RQ2. Is a global $BugLocator$ more effective than local $BugLocator$?
		\end{itemize}
		
		In RQ2, the global corpus is considered both in the direct and indirect relevancy functions to further explore the influence of global data.
		
		\item Heuristic2: Can Word Embedding improve TF.IDF? 
		\begin{itemize}
			\item RQ3. Is a global Doc2Vec better than a global TF.IDF in direct relevancy function?
			
			Similar to RQ1, to evaluate Word Embedding's effect, first we only consider the direct relevancy function and compare the results of Doc2Vec vs. TF.IDF (both are using global data).
			
			\item RQ4. Does a global Doc2Vec improve the effectiveness of a global TF.IDF in indirect relevancy function?
			
			In RQ4, the application of Word Embedding is investigated only in the indirect relevancy function, while the direct relevancy is calculated using a global TF.IDF based function.
			
			\item RQ5. Does combining global Doc2Vec and global TF.IDF improve the effectiveness of global $BugLocator$?
			
			Finally, in RQ5, global Doc2Vec is applied together with global TF.IDF as a complementary algorithm, in order to help TF.IDF both in direct and indirect relevancy functions.
			
		\end{itemize}
	\end{itemize}

	
	
	The above research questions are designed in a way that let us observe the influence of each change individually, and enable further analyzes and comparisons between these approaches, while answering interesting RQs and not blindly compare all possible combinations. 
	
	\subsection{Data Set}
	In a recent study, Jaekwon Lee et al. introduced Bench4BL \cite{Bench4Bl} where they collect source code and bug reports of 51 open source projects including the fix information per bug report, which defines the ground truth for fault localization. The projects consist of 5 commonly used projects in previous IRFL studies as well as 46 new projects with a total of 61,431 Java files and 9,459 bug reports. They report a comprehensive reproduction study of six state-of-the-art IRFL techniques, including $BugLocator$, in order to compare their efficiencies in all 51 projects. Bench4BL is openly accessible to the public, and to the best of our knowledge, is the largest and most up-to-date benchmark data set for IRFL. 
	
	\clearpage
	\begingroup
	\setlength{\LTleft}{-20cm plus -1fill}
	\setlength{\LTright}{-20cm plus -1fill}
		\begin{tabularx}{\textwidth}{|C{.25\textwidth}|C{.15\textwidth}|C{.15\textwidth}|C{.15\textwidth}|C{.15\textwidth}|C{.15\textwidth}|}
			\caption{Details of 51 open-source projects} \label{dataset}
			\\ \hline 
			\rowcolor[HTML]{C0C0C0}{\textbf{Project}} & {\textbf{\# of SC files}} & {\textbf{\# of BR files}}& {\textbf{\# of unique SC words}}&  {\textbf{\# of unique BR words}}& {\textbf{\# of missing BR words}} \\ \hline \endhead
			CSV & 29 & 14 & 987 & 191 & 29 \\ \hline
			MOBILE & 64 & 11 & 1188 & 142 & 27\\ \hline
			ELY & 68 & 7 & 882 & 349 & 113\\ \hline
			WFMP & 80 & 3 & 673 & 193 & 36\\ \hline
			CRYPTO & 82 & 7 & 956 & 98 & 20 \\ \hline
			WEAVER & 113 & 2 & 885 & 34 & 6 \\ \hline
			CODEC & 115 & 42 & 7913 & 833 & 255\\ \hline
			WFARQ & 126 & 1 & 1145 & 53 & 20 \\ \hline
			SHL & 151 & 10 & 1599 & 178 & 27\\ \hline
			SOCIALTW & 153 & 8 & 1394 & 212 & 40\\ \hline
			SOCIALLI & 180 & 4 & 1415 & 45 & 2\\ \hline
			SOCIAL & 212 & 13 & 1598 & 305 & 66\\ \hline
			IO & 227 & 91 & 1379 & 1397 & 449\\ \hline
			BATCHADM & 243 & 20 & 1841 & 453 & 114\\ \hline
			ENTESB & 252 & 16 & 370 & 863 & 304\\ \hline
			SOCIALFB & 253 & 15 & 2282 & 381 & 105\\ \hline
			COMPRESS & 263 & 112 & 2840 & 1896 & 750\\ \hline
			ANDROID & 305 & 9 & 3892 & 236 & 55 \\ \hline
			LANG & 305 & 166 & 5362 & 2735 & 1211\\ \hline
			DATAJPA & 330 & 144 & 2679 & 2475 & 1022\\ \hline
			ZXing & 391 & 20  & 4495 & 776 & 209\\ \hline
			AMQP & 408 & 97 & 2748 & 1811 & 778 \\ \hline
			DATAREST & 414 & 121 & 2518 & 2062 & 809\\ \hline
			CONFIGURATION & 447 & 11 & 3857 & 1578 & 495\\ \hline
			SWT & 484 & 98 & 11418 & 1730 & 677 \\ \hline
			COLLECTIONS & 525 & 39 & 3521 & 893 & 211\\ \hline
			DATAREDIS & 551 & 49 & 3544 & 935 & 295\\ \hline
			LDAP & 566 & 52 & 3937 & 1156 & 295\\ \hline
			DATACMNS & 604 & 152 & 3794 & 2409 & 863\\ \hline
			DATAMONGO & 622 & 264 & 2839 & 3301 & 1481\\ \hline
			SGF & 695 & 98 & 4386 & 1678 & 693\\ \hline
			SECOAUTH & 726 & 66 & 4029 & 1303 & 407\\ \hline
			SWARM & 727 & 54 & 2841 & 1212 & 477\\ \hline
			SWF & 808 & 105 & 4864 & 1924 & 713 \\ \hline
			DATAGRAPH & 848 & 12 & 4286 & 625 & 188\\ \hline
			JBMETA & 858 & 20 & 3524 & 582 & 238\\ \hline
			SWS & 925 & 159 & 4410 & 3065 & 1337 \\ \hline
			SHDP & 1102 & 45 & 4134 & 653 & 169\\ \hline
			ROO & 1109 & 558 & 4870 & 5985 & 2998\\ \hline
			MATH & 1617 & 82 & 5350 & 3391 & 1799\\ \hline
			SEC & 1618 & 362 & 8811 & 5194 & 2021\\ \hline
			BATCH & 1732 & 354 & 8046 & 3296 & 1163\\ \hline
			HBASE & 2714 & 746 & 31360 & 9129 & 4957\\ \hline
			WFCORE & 3532 & 360 & 18939 & 4224 & 1883\\ \hline
			HIVE & 4651 & 1193 & 26640 & 9957 & 5556\\ \hline
			PDE & 5411 & 59 & 16502 & 1272 & 566\\ \hline
			AspectJ & 6485 & 122 & 18630 & 4522 & 2068 \\ \hline
			SPR & 6512 & 123 & 34141 & 2431 & 886\\ \hline
			JDT & 7105 & 94 & 22396 & 2737 & 1370\\ \hline
			WFLY & 8990 & 821 & 22955 & 10110 & 5659 \\ \hline
			CAMEL & 14522 & 1400 & 40514 & 10546 & 5088\\ \hline
		\end{tabularx}
	
	\endgroup

	Therefore, in our study, we use Bench4BL as our global corpus in $Globug$. For each project, we take the latest version of the program as the code corpus of that project and we exclude the bug reports with no fixed files in the considered version. 
	Note that to build the vocabulary, we use the entire benchmark, including the current project, but when we calculate distances the historical global training set is considered as the benchmark (excluding the project under study). This is to ensure we don't have any information leaking in our training phase.
	
	The detail of the data set is summarized in table \ref{dataset}.
	
	To deal with noises in the dataset, we follow the same procedure that has been suggested in $BugLocator$, as follows:
	
	\begin{itemize}
		\item \textbf{Bug Report Pre-processing}: In order to prune the bug reports to improve their quality, first, all stop words are eliminated from them, and then they are all stemmed. 
		\item \textbf{Source Code Pre-processing}: After parsing the Abstract Syntax Tree (AST) of each Java file, the comments, Java keywords, and stop words were removed from each file and each code term was transformed to its stemmed form.
		
	\end{itemize}
		After pre-processing, a dictionary of all source code unique words is created. There are 263,402 unique words in all 51 project source code files, and 53,812 unique words in all bug reports. Since we make the dictionary based on source codes, there will be some words in the bug reports that are missing, which we dismiss them in the TF.IDF vectors. The number of missing bug report words in all projects is 32,345. We dismiss them because they have no information gain for locating related source codes, in an IRFL approach. 
	
	\subsection{Experiment Design}
	
	To answer the RQs, we designed and implemented a set of seven IRFL methods, which are different variations of $BugLocator$ or $GloBug$:
	\begin{itemize}
		\item \textbf{Method 1:} This is essentially $BugLocator$ with only the direct relevancy function (without indirect relevancy function).
		\item \textbf{Method 2:} This is the same as Method 1, except the direct relevancy function is calculated using a global TF.IDF.
		\item \textbf{Method 3:}  $BugLocator$ (using local TF.IDF for both direct and indirect relevancy functions).
		\item \textbf{Method 4:}  $GloBug$ Variation 1 (using global TF.IDF for both direct and indirect relevancy functions).
		\item \textbf{Method 5:}  This is the same as Method 2, except the direct relevancy function is calculated using a global Doc2Vec instead of a global TF.IDF.
		\item \textbf{Method 6:}  $GloBug$ Variation 2 (using global TF.IDF for direct relevancy and global Doc2Vec for indirect relevancy).
		\item \textbf{Method 7:}  This uses both global TF.IDF and global Doc2Vec on both direct and indirect relevancy functions.
	\end{itemize}
	
	In the following, the experiment design per RQ, is explained:
	
	\begin{itemize}
		\item \textbf{RQ1:} In RQ1, the objective is to analyze the effect of the global corpus on the ``direct relevancy'' function, independently. Therefore, we will be using Method 1 and Method 2 for comparison.
		
		\item \textbf{RQ2:} After looking into ``direct relevancy'' function in RQ1, in RQ2, we add the ``indirect relevancy'' function into the picture. So the baseline technique (Method 3) in this RQ is $BugLocator$, which is then compared with Method 4, which is $Globug$.
		
		\item \textbf{RQ3:} In RQ3 to RQ5, the goal is to study the performance of our word embedding technique (Doc2Vec) compared to $TF.IDF$. We explore this in three RQs. In RQ3, we only look at the ``direct relevancy'' function, similar to RQ1, but since Doc2Vec requires a large corpus to train, we only consider Method 2 from RQ1 (using global TF.IDF) as a fair baseline. So we compare Method 2 with Method 5.
		

		
		\item \textbf{RQ4:} In RQ4, we will look at the performance of Doc2vec in the ``indirect relevancy'' function. To implement that we take Method 4 from RQ2 (which is $GloBug$ Variation 1) and compare it with Method 6, which is an implementation of $GloBug$ Variation 2.
		

		
		\item \textbf{RQ5:} Finally, in RQ5 we will look at the performance of a combined technique (Method 7) were we use both global TF.IDF and Global Doc2Vec in both relevancy functions. We will be comparing it with Method 4 (which is $GloBug$ Variation 1).
		
		
		To combine the scores given by TF.IDF and Doc2Vec per relevancy function, we first normalize the score to [0:1] the their average is considered as the final score per relevancy function. 
		
	\end{itemize} 
	\color{black}
	
	\begin{table}[!t]
		\caption{Summary of all seven methods of implementing $BugLocator$ and $GloBug$}
		\label{Methods}
		\resizebox{\textwidth}{!}{\begin{tabular}{cccccccccc}
				\hline
				\cellcolor[HTML]{000000}{\color[HTML]{000000} } & {\color[HTML]{000000} } & \multicolumn{4}{c}{{\color[HTML]{000000} \textbf{Direct Relevancy}}} & \multicolumn{4}{c}{{\color[HTML]{000000} \textbf{Indirect Relevancy}}} \\ \hline
				\multicolumn{1}{c|}{{\color[HTML]{000000} }} & \multicolumn{1}{c|}{{\color[HTML]{000000} }} & \multicolumn{2}{c}{{\color[HTML]{000000} Doc2Vec}} & \multicolumn{2}{c|}{{\color[HTML]{000000} TF.IDF}} & \multicolumn{2}{c}{{\color[HTML]{000000} Doc2Vec}} & \multicolumn{2}{c}{{\color[HTML]{000000} TF.IDF}} \\
				\multicolumn{1}{c|}{{\color[HTML]{000000} }} & \multicolumn{1}{c|}{{\color[HTML]{000000} }} & {\color[HTML]{000000} Local} & {\color[HTML]{000000} Global} & {\color[HTML]{000000} Local} & \multicolumn{1}{c|}{{\color[HTML]{000000} Global}} & {\color[HTML]{000000} Local} & {\color[HTML]{000000} Global} & {\color[HTML]{000000} Local} & {\color[HTML]{000000} Global} \\ \hline
				\rowcolor[HTML]{C0C0C0} 
				\multicolumn{1}{c|}{\cellcolor[HTML]{C0C0C0}{\color[HTML]{000000} }} & \multicolumn{1}{c|}{\cellcolor[HTML]{C0C0C0}{\color[HTML]{000000} \begin{tabular}[c]{@{}c@{}}Method 1\\ Local TF.IDF\end{tabular}}} & {\color[HTML]{000000} -} & {\color[HTML]{000000} -} & {\color[HTML]{000000} \ding{51}} & \multicolumn{1}{c|}{\cellcolor[HTML]{C0C0C0}{\color[HTML]{000000} -}} & {\color[HTML]{000000} -} & {\color[HTML]{000000} -} & {\color[HTML]{000000} -} & {\color[HTML]{000000} -} \\ \cline{2-10} 
				\rowcolor[HTML]{C0C0C0} 
				\multicolumn{1}{c|}{\multirow{-2}{*}{\cellcolor[HTML]{C0C0C0}{\color[HTML]{000000} \textbf{RQ1}}}} & \multicolumn{1}{c|}{\cellcolor[HTML]{C0C0C0}{\color[HTML]{000000} \begin{tabular}[c]{@{}c@{}}Method 2\\ Global TF.IDF\end{tabular}}} & {\color[HTML]{000000} -} & {\color[HTML]{000000} -} & {\color[HTML]{000000} -} & \multicolumn{1}{c|}{\cellcolor[HTML]{C0C0C0}{\color[HTML]{000000} \ding{51}}} & {\color[HTML]{000000} -} & {\color[HTML]{000000} -} & {\color[HTML]{000000} -} & {\color[HTML]{000000} -} \\ \hline
				\multicolumn{1}{c|}{{\color[HTML]{000000} }} & \multicolumn{1}{c|}{{\color[HTML]{000000} \begin{tabular}[c]{@{}c@{}}Method 3 -\\ BugLocator\end{tabular}}} & {\color[HTML]{000000} -} & {\color[HTML]{000000} -} & {\color[HTML]{000000} \ding{51}} & \multicolumn{1}{c|}{{\color[HTML]{000000} -}} & {\color[HTML]{000000} -} & {\color[HTML]{000000} -} & {\color[HTML]{000000} \ding{51}} & {\color[HTML]{000000} -} \\ \cline{2-10} 
				\multicolumn{1}{c|}{\multirow{-2}{*}{{\color[HTML]{000000} \textbf{RQ2}}}} & \multicolumn{1}{c|}{{\color[HTML]{000000} \begin{tabular}[c]{@{}c@{}}Method 4 -\\ GloBug Variation 1\end{tabular}}} & {\color[HTML]{000000} -} & {\color[HTML]{000000} -} & {\color[HTML]{000000} -} & \multicolumn{1}{c|}{{\color[HTML]{000000} \ding{51}}} & {\color[HTML]{000000} -} & {\color[HTML]{000000} -} & {\color[HTML]{000000} -} & {\color[HTML]{000000} \ding{51}} \\ \hline
				\rowcolor[HTML]{C0C0C0} 
				\multicolumn{1}{c|}{\cellcolor[HTML]{C0C0C0}{\color[HTML]{000000} }} & \multicolumn{1}{c|}{\cellcolor[HTML]{C0C0C0}{\color[HTML]{000000} \begin{tabular}[c]{@{}c@{}}Method 2 -\\ Global TF.IDF\end{tabular}}} & {\color[HTML]{000000} -} & {\color[HTML]{000000} -} & {\color[HTML]{000000} -} & \multicolumn{1}{c|}{\cellcolor[HTML]{C0C0C0}{\color[HTML]{000000} \ding{51}}} & {\color[HTML]{000000} -} & {\color[HTML]{000000} -} & {\color[HTML]{000000} -} & {\color[HTML]{000000} -} \\ \cline{2-10} 
				\rowcolor[HTML]{C0C0C0} 
				\multicolumn{1}{c|}{\multirow{-2}{*}{\cellcolor[HTML]{C0C0C0}{\color[HTML]{000000} \textbf{RQ3}}}} & \multicolumn{1}{c|}{\cellcolor[HTML]{C0C0C0}{\color[HTML]{000000} \begin{tabular}[c]{@{}c@{}}Method 5 -\\ Global Doc2Vec\end{tabular}}} & {\color[HTML]{000000} -} & {\color[HTML]{000000} \ding{51}} & {\color[HTML]{000000} -} & \multicolumn{1}{c|}{\cellcolor[HTML]{C0C0C0}{\color[HTML]{000000} -}} & {\color[HTML]{000000} -} & {\color[HTML]{000000} -} & {\color[HTML]{000000} -} & {\color[HTML]{000000} -} \\ \hline
				\multicolumn{1}{c|}{{\color[HTML]{000000} }} & \multicolumn{1}{c|}{{\color[HTML]{000000} \begin{tabular}[c]{@{}c@{}}Method 4 -\\ GloBug Variation 1\end{tabular}}} & {\color[HTML]{000000} -} & {\color[HTML]{000000} -} & {\color[HTML]{000000} -} & \multicolumn{1}{c|}{{\color[HTML]{000000} \ding{51}}} & {\color[HTML]{000000} -} & {\color[HTML]{000000} -} & {\color[HTML]{000000} -} & {\color[HTML]{000000} \ding{51}} \\ \cline{2-10} 
				\multicolumn{1}{c|}{\multirow{-2}{*}{{\color[HTML]{000000} \textbf{RQ4}}}} & \multicolumn{1}{c|}{{\color[HTML]{000000} \begin{tabular}[c]{@{}c@{}}Method 6 - GloBug Variation 2\end{tabular}}} & {\color[HTML]{000000} -} & {\color[HTML]{000000} -} & {\color[HTML]{000000} -} & \multicolumn{1}{c|}{{\color[HTML]{000000} \ding{51}}} & {\color[HTML]{000000} -} & {\color[HTML]{000000} \ding{51}} & {\color[HTML]{000000} -} & {\color[HTML]{000000} -} \\ \hline
				\rowcolor[HTML]{C0C0C0} 
				\multicolumn{1}{c|}{\cellcolor[HTML]{C0C0C0}{\color[HTML]{000000} }} & \multicolumn{1}{c|}{\cellcolor[HTML]{C0C0C0}{\color[HTML]{000000} \begin{tabular}[c]{@{}c@{}}Method 4 - \\ GloBug Variation 1\end{tabular}}} & {\color[HTML]{000000} -} & {\color[HTML]{000000} -} & {\color[HTML]{000000} -} & \multicolumn{1}{c|}{\cellcolor[HTML]{C0C0C0}{\color[HTML]{000000} \ding{51}}} & {\color[HTML]{000000} -} & {\color[HTML]{000000} -} & {\color[HTML]{000000} -} & {\color[HTML]{000000} \ding{51}} \\ \cline{2-10} 
				\rowcolor[HTML]{C0C0C0} 
				\multicolumn{1}{c|}{\multirow{-2}{*}{\cellcolor[HTML]{C0C0C0}{\color[HTML]{000000} \textbf{RQ5}}}} & \multicolumn{1}{c|}{\cellcolor[HTML]{C0C0C0}{\color[HTML]{000000} \begin{tabular}[c]{@{}c@{}}Method 7 - Combined\end{tabular}}} & {\color[HTML]{000000} -} & {\color[HTML]{000000} \ding{51}} & {\color[HTML]{000000} -} & \multicolumn{1}{c|}{\cellcolor[HTML]{C0C0C0}{\color[HTML]{000000} \ding{51}}} & {\color[HTML]{000000} -} & {\color[HTML]{000000} \ding{51}} & {\color[HTML]{000000} -} & {\color[HTML]{000000} \ding{51}} \\ \hline
		\end{tabular}}
	\end{table}
	
	The explained mapping between all seven methods and the RQs they answer is represented in table \ref{Methods}. 
		\subsection{Doc2Vec Parameters}
		In general there are several Doc2Vec implementations in the literature. The major ones are Paragraph Vector, using a distributed memory model (PV-DM), and Paragraph Vector using Distributed bag of words (PV-DBOW). We have employed a model that uses combination of the two methods (similar to the Doc2Vec's original paper). We have used the following default settings in our experiments. For the PV-DM model: vector-size = 100, alpha = 0.045, window-size = 5, min-count = 2, and min-alpha = alpha/2. For PV-DBOW model: the vector-size and alpha are the same as PV-DM, the negative parameter is 5, hs = 0, min-count = 2, sample = 0, and min-alpha = alpha/3.
	
	\subsection{Evaluation Metrics}
	
	To analyze the effectiveness of each method, most IRFL studies use Mean Reciprocal Rank (MRR) and Mean Average Precision (MAP). We also use MRR and MAP as two main metrics. In addition, we also report the Top N Rank results. 
	
	In each comparison, we run Wilcoxon Signed Rank test\cite{Wilcoxon}, which is a non-parametric paired hypothesis test, and we report the P-value to investigate any statistically significant difference in the results.
	
	\subsubsection{MRR}
	MRR is a statistical metric for evaluating an IR method that produces a list of possible responses to a query. The reciprocal rank of a query is the multiplicative inverse of the rank of the first correct answer. The mean reciprocal rank is the average of the reciprocal ranks of results of a set of queries Q, and is calculated as follows:
	
	\begin{equation}
	MRR = \frac{1}{\left | Q \right |}\sum_{i=1}^{\left | Q \right |}\frac{1}{rank_{i}} 
	\end{equation}
	
	Therefore, MRR indicates the average rank of the first correctly retrieved buggy file in the predicted list of source code files, when localizing the fault for a set of query bug reports (Q). Therefore, the higher the MRR value, the better the bug localization performance.
	
	\subsubsection{MAP}
	
	MAP is another metric for measuring the quality of an IR method, when a query may have multiple relevant documents. The Average Precision of a single query (AvgP) is the average of the precision values obtained for the query, which is computed as follows:
	
	\begin{equation}
	avgP_{i}=\sum_{1}^{M}\frac{p(j)\times pos(j)}{number\:of\:positive\:instances}
	\end{equation}
	
	In the equation above, j is the rank, M is the number of instances retrieved, pos(j) indicates whether the instance in the rank j is relevant or not. P(j) is the precision at the given cut-off rank j, and is defined as follows:
	
	\begin{equation}
	P(j)=\frac{number\:of\:possitive\:instances\:in\:top\:j\:possitions}{j}
	\end{equation}
	
	Then, the MAP for a set of queries is the mean of the average precision values for all queries. The higher the MAP value, the better the bug localization performance.
	
	\subsubsection{Top N Rank}
	
	A study by Parnin et. al. \cite{TopK} shows that FL techniques are helpful only when the root causes are ranked at a high absolute position. Therefore, we further use $Top N Rank$ to measure the efficiency of an IRFL method. 
	
	Top N Rank is the number of bugs whose associated files are ranked in the top N (e.g. N= 1, 5, 10 in this study) of the returned results. Given a bug report, if the top N query results contain at least one file at which the bug should be fixed, we consider the bug as located. The higher the metric value, the better the bug localization performance.
	
	\subsection{Experiment Results}
	\label{Results_section}
	
	In this section, first, the results of all seven techniques are represented to answer the five research questions. Next, we further discuss our two heuristics to see how successful they have been in our studied IRFL techniques.

	\subsubsection{Answers to RQs}
	In order to answer each research question, the performance of its two associated methods (based on Table \ref{Methods}) is visualized and compared against one another in terms of MAP and MRR as two separate bar-charts. The 51 projects are sorted based on the size of their source code repository (number of Java files) and the size of their bug report repository (number of bug reports). Note that in the comparisons, any difference less than 2\% is considered as insignificant difference.

	\textbf{Answer to RQ1: ``Can a global corpus improve effectiveness of the direct relevancy function in $BugLocator$?''}
	
	\begin{figure}
		\begin{turn}{-90}
			\begin{minipage}{1.25\linewidth}
				\includegraphics[width=\linewidth]{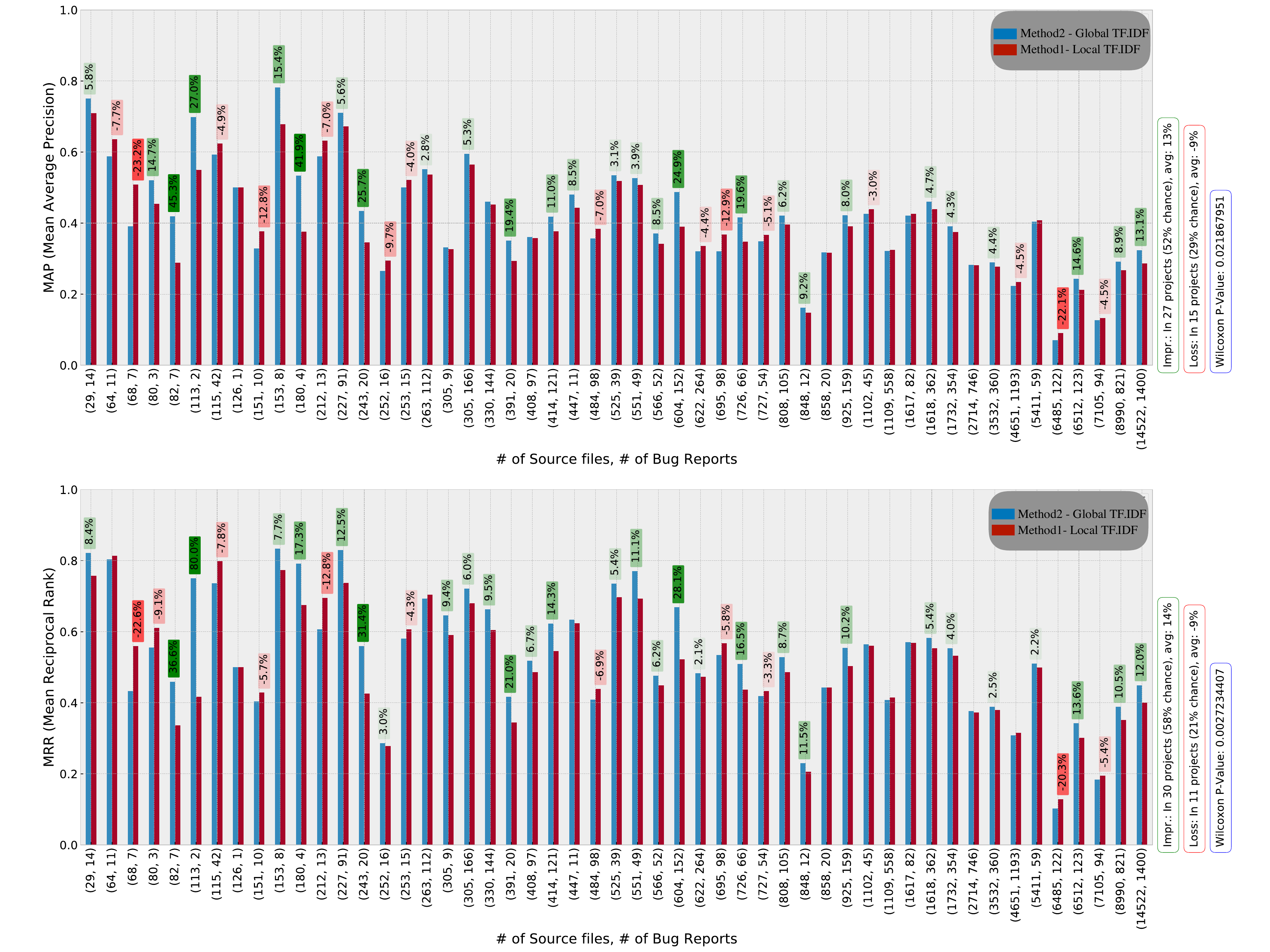}
			\end{minipage}
		\end{turn}
		\caption{RQ1 results in terms of MRR and MAP over 51 projects}
		\label{RQ1 Results}
	\end{figure}
	
	
	Figure \ref{RQ1 Results} represents the performance of Method 1 (Local TF.IDF) and Method 2(Global TF.IDF), where both only use a direct relevancy function. As we see, using the global corpus in the direct relevancy function, on average, causes 14\% improvement in 30 projects (58\% of projects) in terms of MRR, while it improves Local TF.IDF, on average 13\% in terms of MAP in 27 projects (52\% of projects).
	
	However, in some projects, the global corpus cannot contribute to the accuracy of Method 1 (Local TF.IDF). We see that in 21\% of projects (11 projects) with an average loss of 9\% in MRR, and in 29\% of cases (15 projects) with an average of 9\% loss in MAP, global corpus (Method 2) fails to improve the Method 1 (Local TF.IDF). There is however, no clear evidence on the correlation between the size of the project or its \# of bug reports with the performance of Method 2 (Global TF.IDF).
	
	To assure that the difference in the results is not due to the chance, we performed Wilcoxon signed rank test with H0=``The median results of two techniques are not statistically different''. The results indicate that with a significance level of $\alpha=0.05$ , we can reject the null hypothesis both in the MRR (p-value$=0.0027 < 0.05$) and MAP (p-value=$0.02187 < 0.05$).  
	
	\begin{figure}
		\begin{turn}{-90}
			\begin{minipage}{1.25\linewidth}
				\includegraphics[width=\linewidth]{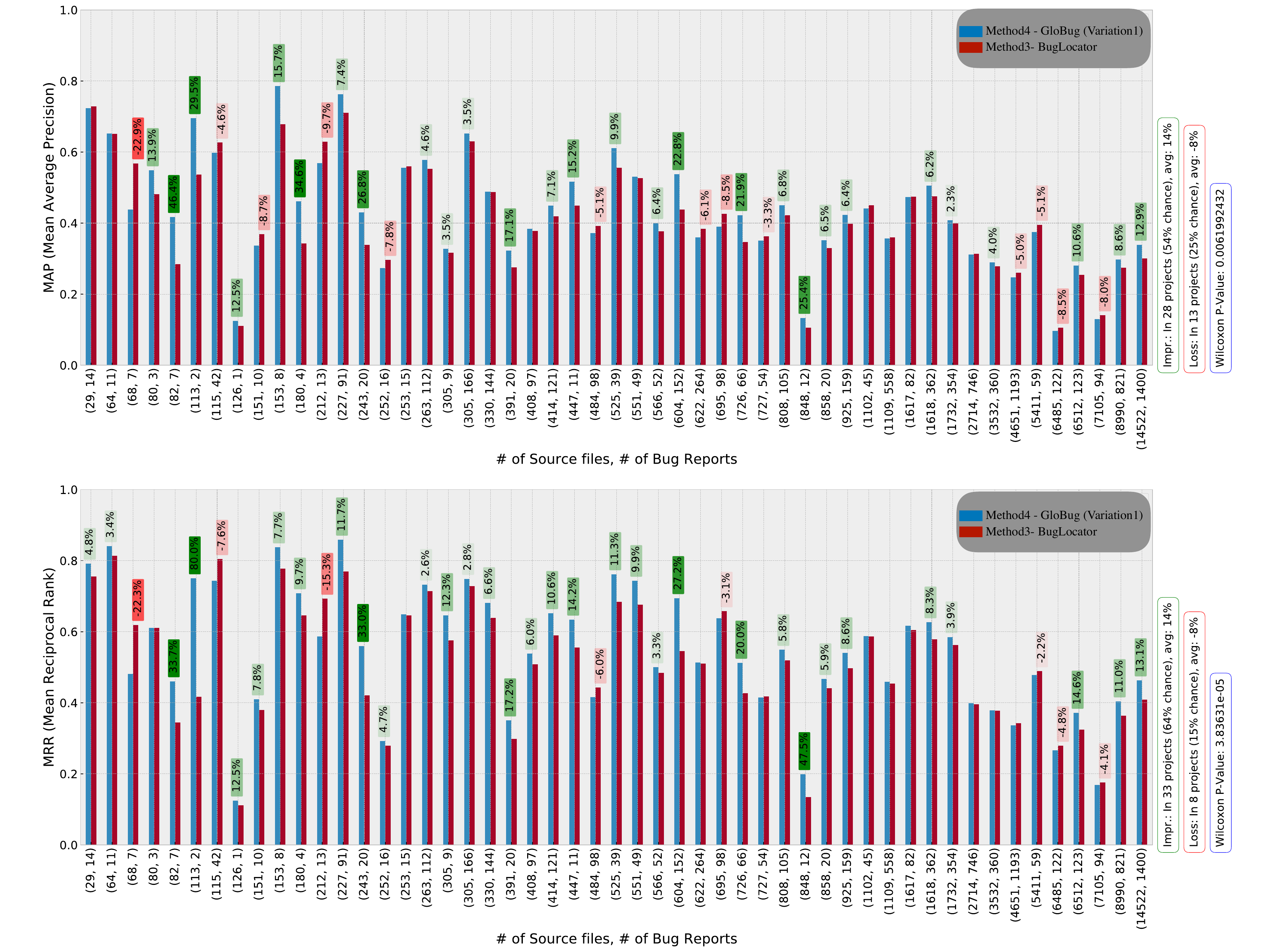}
			\end{minipage}
		\end{turn}
		\caption{RQ2 results in terms of MRR and MAP, over 51 projects}
		\label{RQ2 Results}
	\end{figure}
	
	Based on these results, we answer RQ1 as:
	
	\textbf{Answer: }
	``Yes, in most cases, global direct relevancy outperforms the local relevancy, while using TF.IDF.'' 

	\textbf{Answer to RQ2: ``Is a Global $BugLocator$ more effective than local $BugLocator$?''}
	
	
	Figure \ref{RQ1 Results} represents the performance of original $BugLocator$ (Method 3) and a Global $BugLocator$ (Method 4: $GloBug$-Variation1) in terms of MAP and MRR as two separate bar-charts.
	
	As we see, Method 4 outperforms $BugLocator$ in 33 projects (64\% of projects) with an average of 14\% in terms of MRR, while it improves MAP with an average of 14\% in 28 projects (54\% of projects).
	
	On the other hand, in 15\% of the cases (8 projects), there is a chance of 8\% loss in MRR and 25\% (13 projects ) chance of 8\% loss in MAP, on average. We again, do not see any correlation between size of project or number of bug reports and the performance of the IRFL methods.  
	
	The result of Wilcoxon test indicates that with a significance level of $\alpha=0.05$ , we can reject the null hypothesis both in the MRR (p-value$=0.0< 0.05$) and MAP (p-value=$0.0062 < 0.05$). Meaning that the difference between the results of the two methods is not statistically insignificant. 
	
	Hence, RQ2 is answered as:
	
	\textbf{Answer: }
	``Yes, a Global $BugLocator$ (i.e., $GloBug$-Variation1) outperforms the original local $BugLocator$ in most cases and when it does not, the loss is small."
	
	\textbf{Answer to RQ3 ``Is global Doc2Vec better than global TF.IDF, in direct relevancy function?''}
	
	\begin{figure}
		\begin{turn}{-90}
			\begin{minipage}{1.25\linewidth}
				\includegraphics[width=\linewidth]{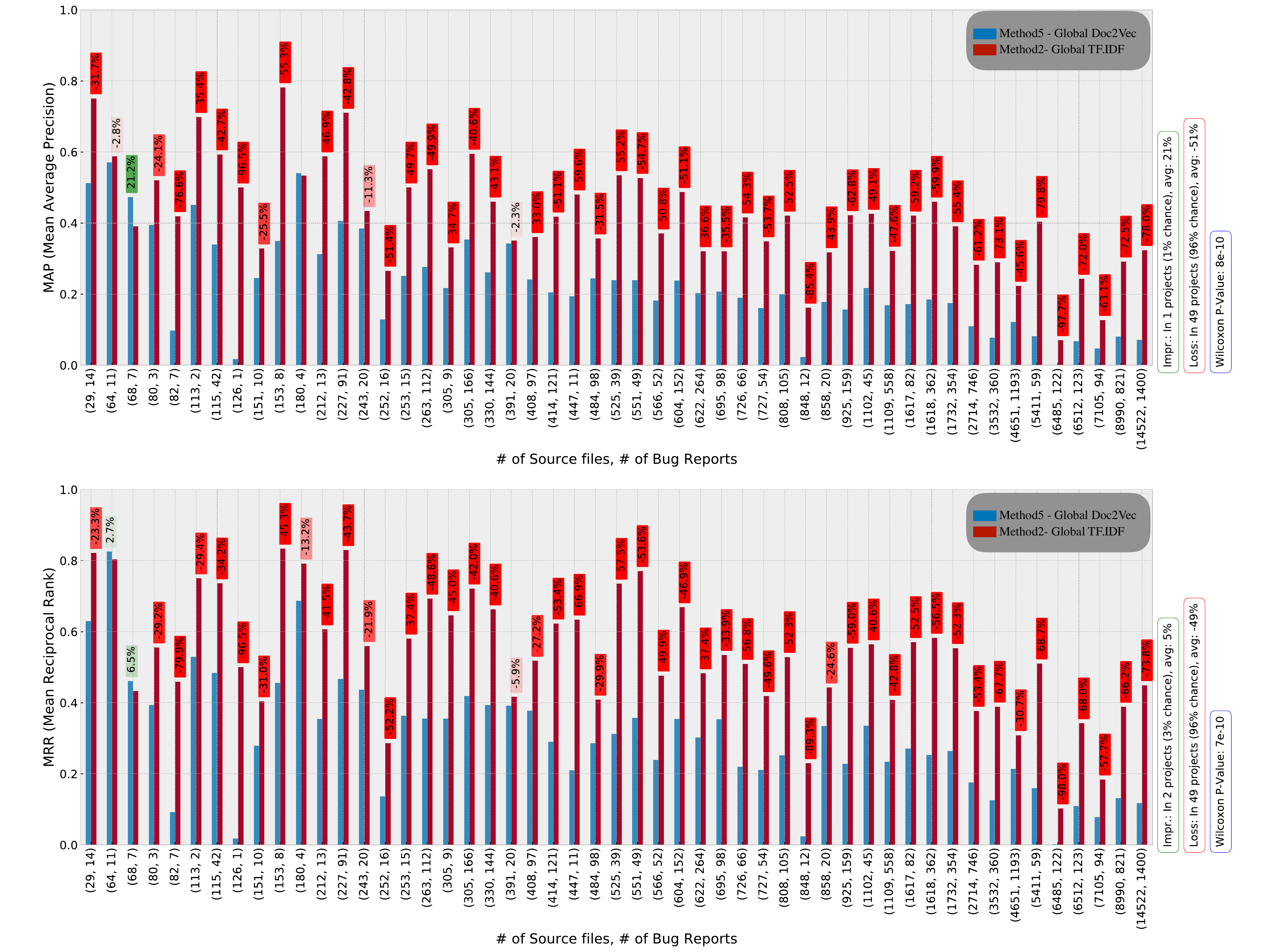}
			\end{minipage}
		\end{turn}
		\caption{RQ3 results in terms of MRR and MAP over 51 projects}
		\label{RQ3 Results}
	\end{figure}
	
	
	The RQ3 results are represented in Figure \ref{RQ3 Results} in terms of MAP and MRR. As we see, when Global TF.IDF in Method 2 is replaced by a Global Doc2Vec in direct relevancy function (Method 5), the efficiency of fault localization drops drastically in 49 projects (96\% of the projects) with an average loss of 49\% in MRR and an average loss of 51\% in MAP. Only, in a rare case (1 project), we observe an improvement of 6.5\% and 21.2\% in terms of MRR and MAP, respectively. 
	
	The Wilcoxon test result also indicates that with a significance level of $\alpha=0.05$ , we can strongly reject the null hypothesis and conclude that there is a statistically significant difference both in the MRR (p-value$=0.0< 0.05$) and MAP (p-value=$0.0 < 0.05$) in the performances of the two analyzed techniques.  
	
	Therefore, RQ3 is answered as:
	
	\textbf{Answer: }``Having Global Doc2Vec as a stand-alone direct relevancy function, is not a good choice, and it may cause a drastic efficiency drop in most cases.''
	
	\textbf{Answer to RQ4 ``Does global Doc2Vec improve the effectiveness of global TF.IDF in indirect relevancy function?"}
	
	\begin{figure}
		\begin{turn}{-90}
			\begin{minipage}{1.25\linewidth}
				\includegraphics[width=\linewidth]{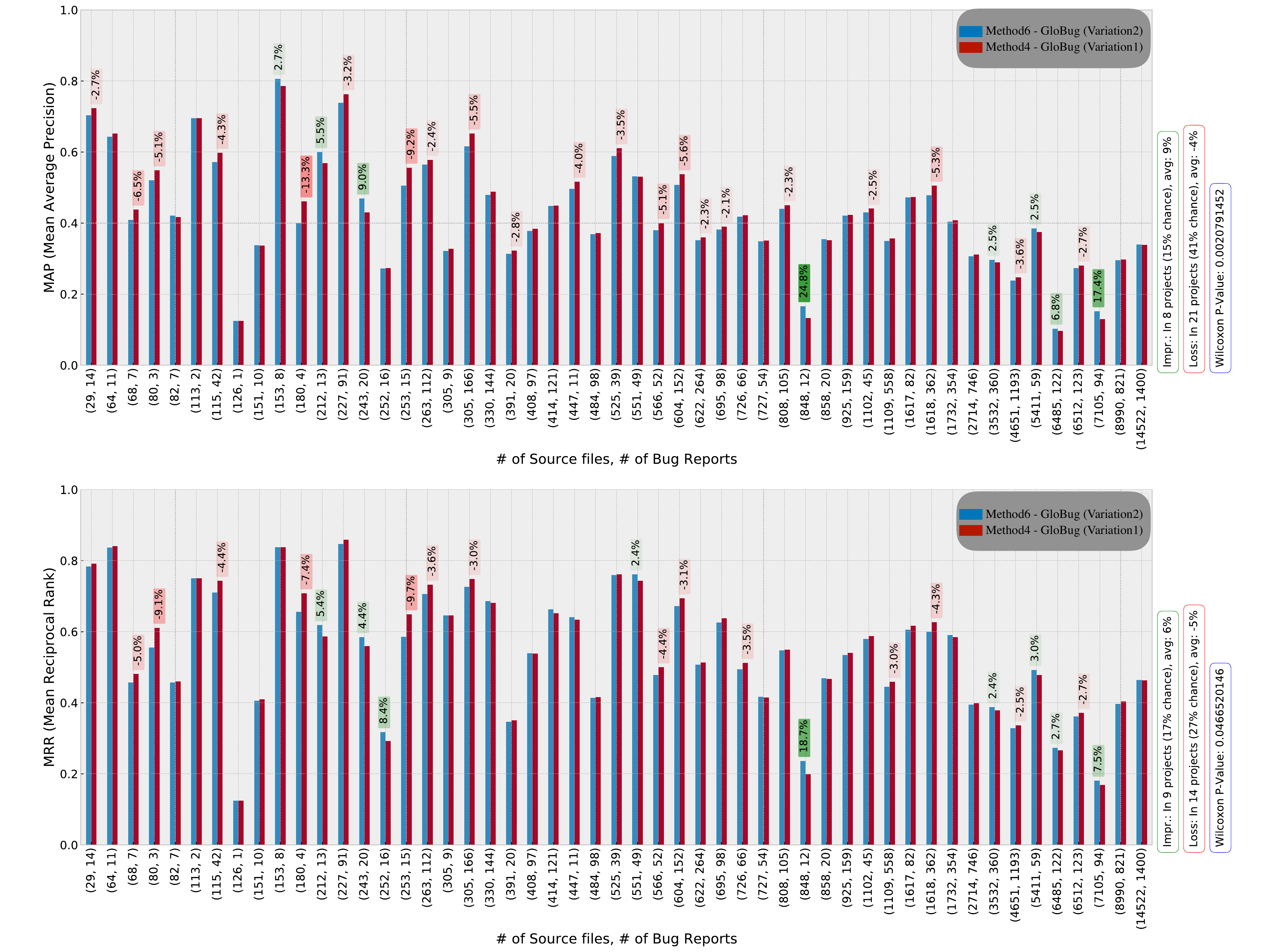}
			\end{minipage}
		\end{turn}
		\caption{RQ4 results in terms of MRR and MAP over 51 projects}
		\label{RQ4 Results}
	\end{figure}
	
	Figure \ref{RQ4 Results} represents the comparison between $GloBug$-Variation 1 (Method 4) and Variation 2 (Method 6), in terms of MRR and MAP. 
	As we see, replacing global TF.IDF-based indirect relevancy function with a global Doc2Vec results in an average improvement of 6\% in MRR in almost 17\% of the projects (9 projects), and an average of 9\% improvement in MAP in 15\% of the projects (8 projects).
	
	However, in other 27\% (14 projects) and 41\% (21 projects) of the cases, with average of 5\% and 4\% loss in MRR and MAP, $GloBug$-Variation 2 fails to perform as good as $GloBug$-Variation 1.
	
	Although the Wilcoxon test results indicates significant difference (with significance level of $\alpha=0.05$), both in terms of MRR (p-value$=0.0467< 0.05$) and MAP (p-value=$0.00208 < 0.05$), looking at the actual values, the two techniques seem to be practically very close in terms of MRR and MAP,
	
	Therefore, the answer to RQ4 is: 
	
	\textbf{Answer: } ``No, $GloBug$-Variation 2 indeed performs worse (Avr: 5\% and 4\% loss in MRR and MAP) than $GloBug$-Variation 1, in most cases. However, in cases where it is better, the improvements are higher (Avr: 6\% and 9\% in MRR and MAP). Therefore, more careful studies are needed when applying Doc2Vec in a IRFL.

	\textbf{Answer to RQ5: ``Does combining global Doc2Vec and global TF.IDF improve the effectiveness of global $BugLocator$?"}

	\begin{figure}
		\begin{turn}{-90}
			\begin{minipage}{1.25\linewidth}
				\includegraphics[width=\linewidth]{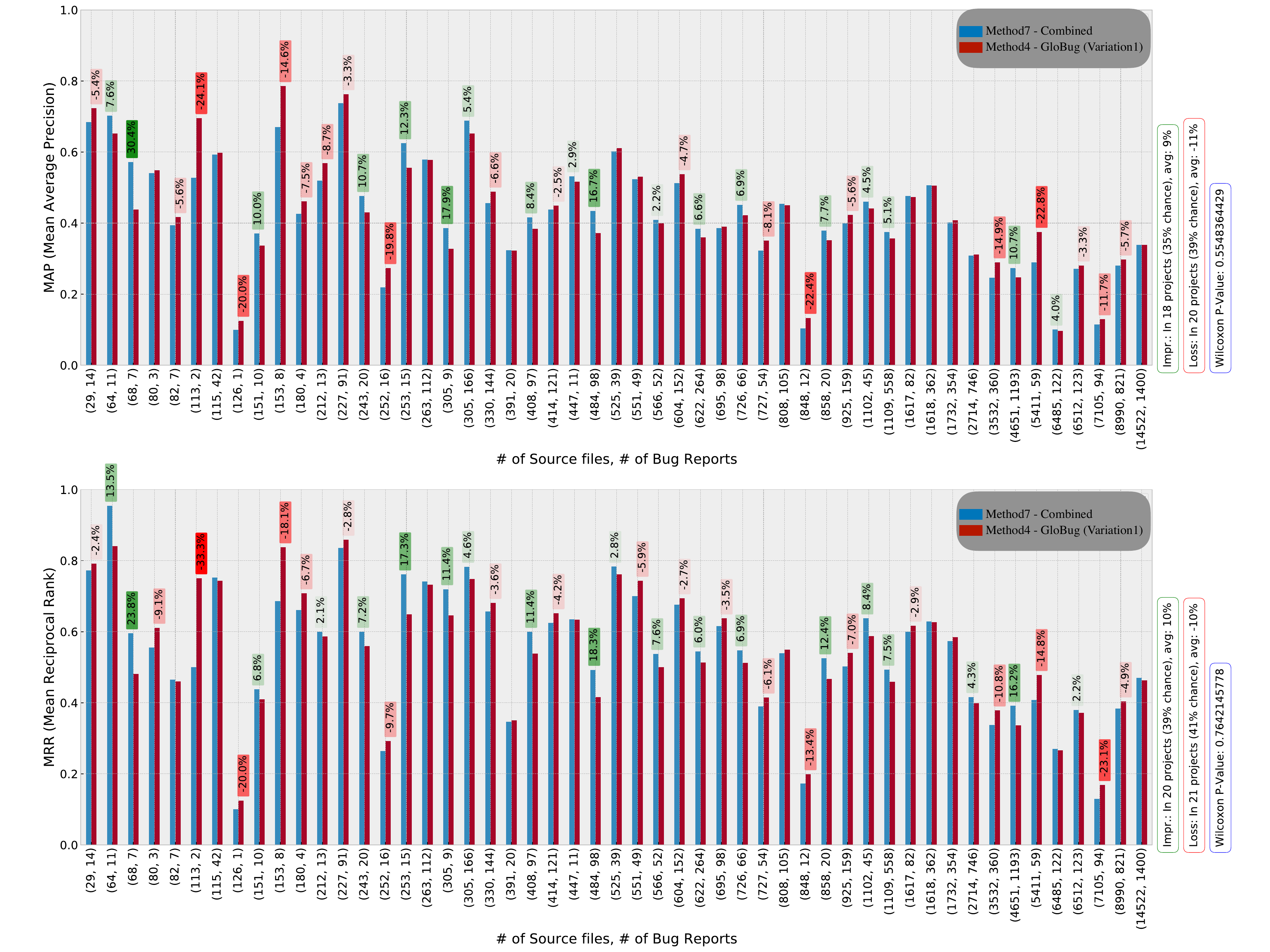}
			\end{minipage}
		\end{turn}
		\caption{RQ5 results in terms of MRR and MAP over 51 projects}
		\label{RQ5 Results}
	\end{figure}
	
	As explained in RQ3 and RQ4, a global Doc2Vec model may not always be a good substitution for the global TF.IDF in the direct or indirect relevancy functions, and in some cases it may even cause a drastic drop in the FL performance. Therefore, in RQ5, the goal is to assess the performance of a combined method (Method 7) in which both direct and indirect relevancy functions are implemented using a combination (average score) of global Doc2vec and global TF.IDF.
	
	Figure \ref{RQ5 Results} represents the performance of ``Combined'' method (Method 7) and $GloBug$-Variation 1 (Method 4), comparing their MRR and MAP, in 51 projects. As we see, in 39\%of projects (20 projects) the combined method outperforms the Method 4, with an average of 10\% improvement in terms of MRR. However, in other 41\% of cases (21 projects), it causes an average loss of 10\%. 
	
	\begin{figure}
		\begin{turn}{-90}
			\begin{minipage}{1.25\linewidth}
				\includegraphics[width=\linewidth]{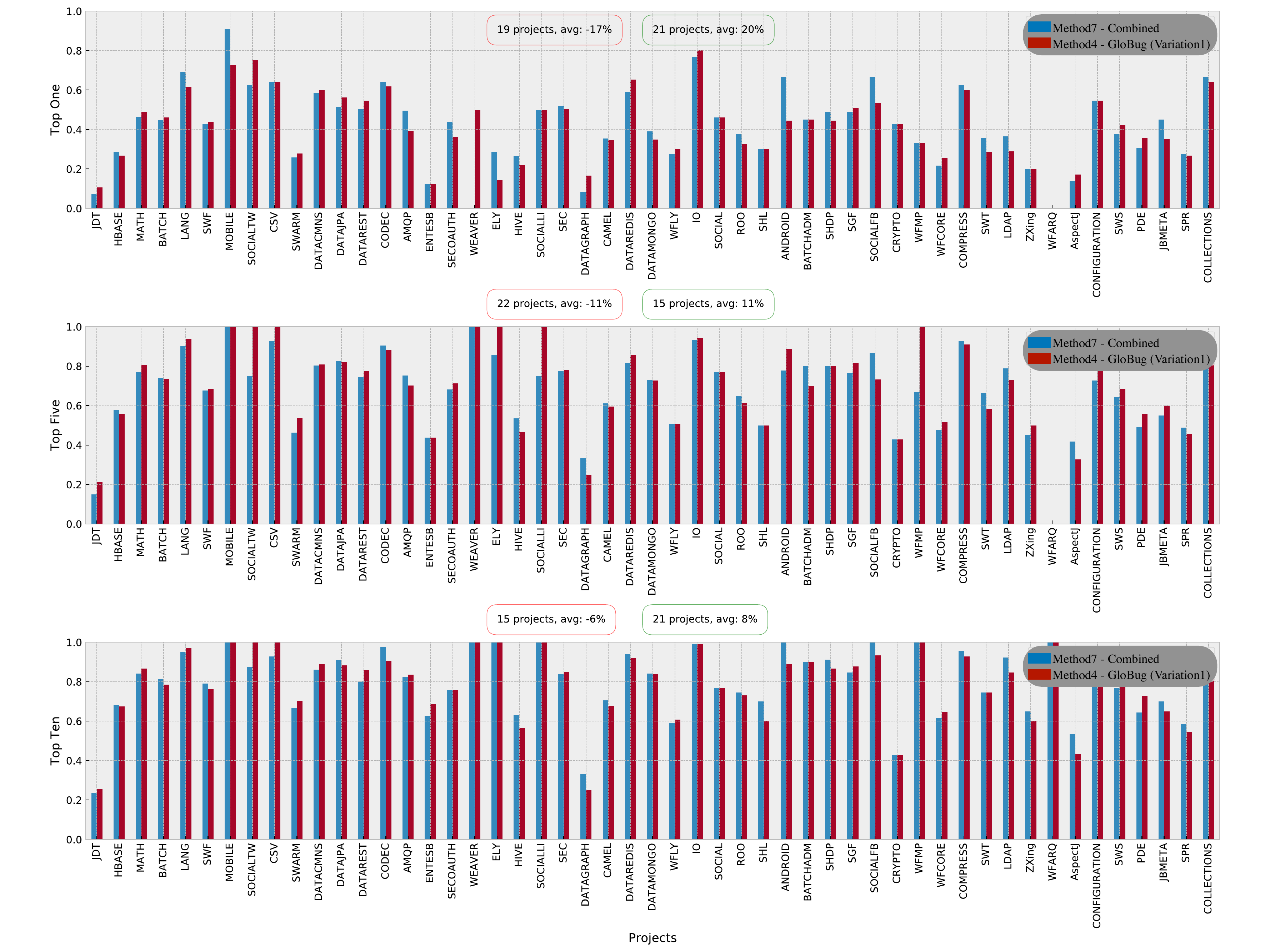}
			\end{minipage}
		\end{turn}
		\caption{RQ5 results in terms of Top N Ranks (N=1,5,10) over 51 projects}
		\label{RQ5_tops_bar}
	\end{figure}
	
	We also see that in terms of MAP, in 35\% of projects (18 projects), the ``Combined'' method causes an average of 9\% improvement, while in other 39\% of studied projects (20 projects), it results in an average loss of 11\%. 
	In RQ5, also we performed Wilcoxon signed rank test. However, this time the results indicate that we fail to reject the null hypothesis at the significance level of $\alpha=0.05$ both in the MRR (p-value$=0.7642 > 0.05$) and MAP (p-value=$0.55484 > 0.05$). This means that there is no statistically significant difference in the performance of the two analyzed methods.
	Since the performance results of combined method and Method 4 are very close in terms of MRR and MAP, we further analyzed the performance of the two techniques by conducting a deeper study in which the two methods are monitored in terms of Top N Ranks.

	
	Figure \ref{RQ5_tops_bar}, represents the performance of the two methods in retrieving the first relevant file in the top one, five and ten ranked files, as three figures. We see that the ``Combined'' method is performing better than Method 4 in 41\% of cases (21 projects) in ranking the relevant files at the top one with 20\% improvement on average. However, in 19 projects (37\% of cases), the number of first ranked buggy files is less in the ``Combined'' method by 17\%, compared to Method 4, on average. Therefore, we conclude that the ``Combined'' method is better at retrieving the buggy files as its first ranked file compared to Method 4 ($GloBug$-Variation 1), in the studied 51 projects. 
	
	Observing the next two charts, we see that Method 4 is slightly better than the ``Combined'' method in ranking the buggy files in the top five ranked files. However, the ``Combined'' method is better in retrieving the buggy files in the top ten ranked files. 
	
	Now, considering all the aforementioned analyzes, we summarize RQ5 as:
	
	\textbf{Answer: } 
	Although Global Doc2Vec is contributing in capturing the semantics of documents in direct and indirect relevancy functions, in some projects, by adding too much complexity to the TF.IDF, it decreases the accuracy of the FL process. Therefore, application of Doc2Vec as an IRFL method must be accompanied with cautious.

	\section{Discussion}
	\label{discussion_section}
	
	This study was formed around two heuristics to analyze the effect of global data and a complex Word Embedding technique on current IRFL methods. In the following two subsections, we further explain the findings with respect to each heuristic.
	
	\subsection{Discussion from the perspective of the first heuristic: Global Data}
	
	
	\begin{figure}[!t]
		\includegraphics[width=\linewidth]{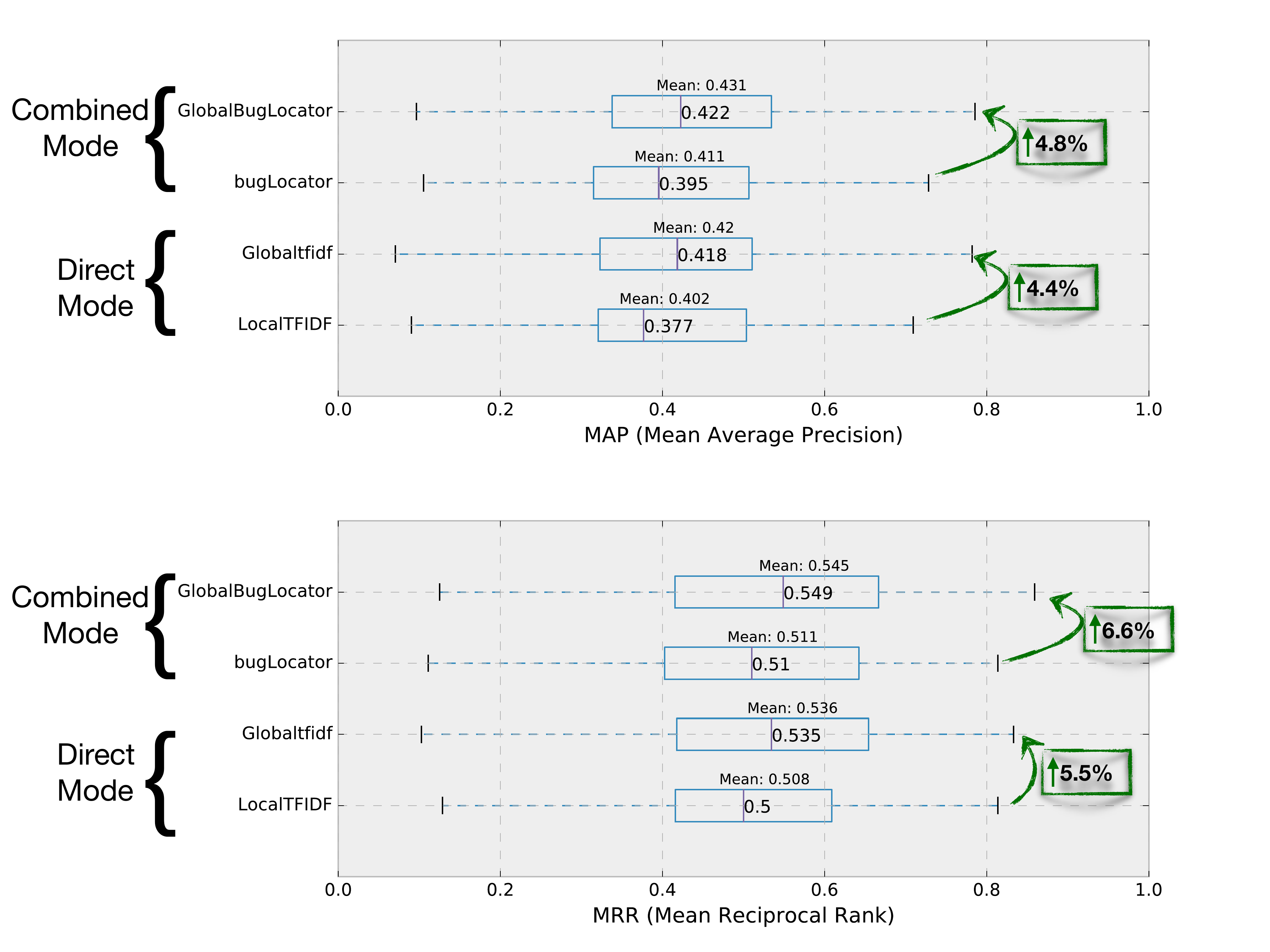}
		\caption{Performance of our first heuristic methods on the 51 projects}
		\label{Our Firs Heuristic - Boxplots}
	\end{figure}
	
	Figure \ref{Our Firs Heuristic - Boxplots} summarizes the performance (in terms of MAP and MRR) distribution of the four methods evaluated in heuristic 1 (RQ1 and RQ2), across the 51 projects. Comparing the mean performance of the first two methods, we see that global training will cause an average improvement of 4.4\% (from 0.402 to 0.42) in terms of MAP. Also, in terms of MRR, we see an average improvement of 5.5\% (from 0.508 to 0.536), over the 51 projects.

	Comparing the mean performance of Method 3 and Method 4, we also see an average improvement of 4.8\% (from 0.411 to 0.431), in terms of MAP, and 6.6\% (from 0.511 to 0.545), in terms of MRR, in the 51 projects. This amount of improvement rates are significant compared to the improvement rates of other IRFL techniques that were proposed after $BugLocator$. 
	
	In this section we take five state-of-the-art $BugLocator$ extensions that are studied in the Bench4Bl paper \cite{Bench4Bl} (BLUiR \cite{BLUiR}, BRTracer \cite{BRTracer}, Amalgam \cite{Amalgam}, BLIA \cite{BLIA}, and Locus \cite{Locus}) and compare their improvements over $BugLocator$ to our approaches' improvements. 
	
	Note that these methods do not make any changes to $BugLocator$'s relevancy functions. They only add new features to $BugLocator$ by exploiting unused information existing in the software projects (see the related work section, Section \ref{RW_section}, for more details). In other words, in theory Global data can be introduced to all these tools as well.

	\begin{figure}[!t]
		\includegraphics[width=\linewidth]{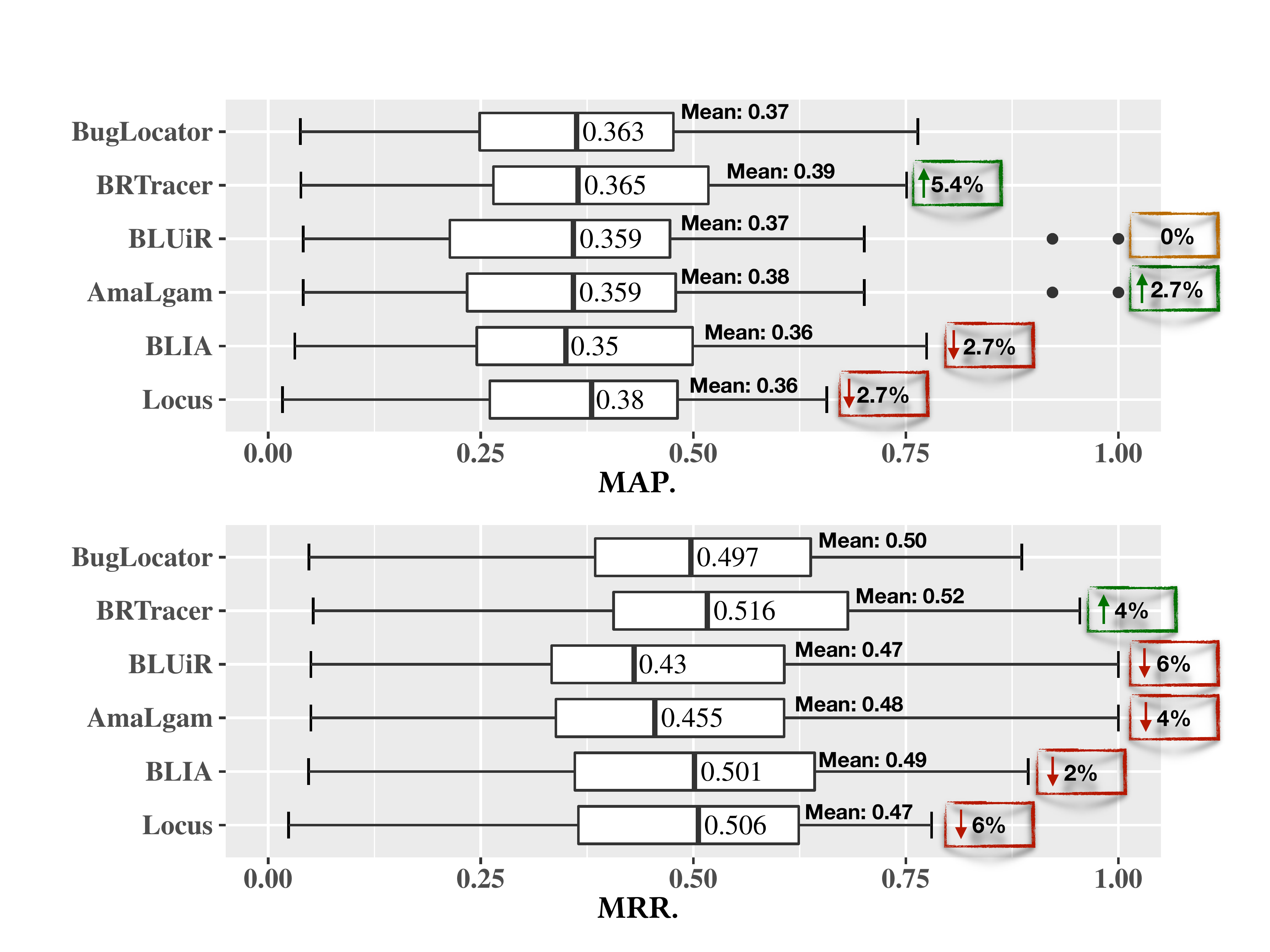}
		\caption{Performance of five state-of-the-art IRFL techniques along with $BugLocator$ on the 51 projects. Data are collected from Lee et at. study \cite{Bench4Bl}.}
		\label{Previous Methods}
	\end{figure}

	Figure \ref{Previous Methods} shows the performance distribution of the five recent IRFL tools along with the $BugLocator$'s. Comparing the mean performance of the five methods with $BugLocator$\footnote{Note that the performance mismatching of the $BugLocator$ between Figure \ref{Our Firs Heuristic - Boxplots} and Figure \ref{Previous Methods} is due to the versioning differences of the subject projects between our study and Bench4BL (we used the most up-to-date versions). However, this will not affect our relative comparisons. We contacted the authors of Bench4BL and confirmed this with them.}, we see that BLIA and Locus not only fail to contribute to the average performance of $BugLocator$ but also with an average loss of 2.7\% in terms of MAP and average losses of 2\% and 6\% in terms of MRR, decrease the efficiency of $BugLocator$. Although Amalgam improves the mean performance of $BugLocator$ in terms of MAP, with an average loss of 4\% it reduces the effectiveness of $BugLocator$ in terms of MRR. The mean performance of BLUiR is almost identical to the mean performance of $BugLocator$ in terms of MAP. However, its efficiency is less than $BugLocator$ by a rate of 4\% in terms of MRR. 
	
	Among these five IRFL methods, BRTracer is showing the best mean performance. BRTracer leverages the stack trace data provided in the bug reports to improve the effectiveness of the relevancy functions in $BugLocator$. Applied on the 51 projects, BRTracer improves the mean performance of the $BugLocator$ by average rates of 5.4\% and 4\% in terms of MAP and MRR, respectively. Although the mean improvement of BRTracer is significant compared to the other methods, this improvement is not reflected in the median values. We see that in terms of median values, BRTracer improves $BugLocator$ only by small rates of 0.5\% (0.363 to 0.365) and 3.8\% (0.497 to 0.516) in terms of MAP and MRR. 
	
	$GloBug$ without extracting any new data from the software project (like stack trace) and only using the available global training data (Bench4Bl) once offline, not only improves the mean performance of $BugLocator$ significantly (4.8\% and 6.6\% in terms of MAP and MRR), but also improves the median values by rates of 6.8\% (0.395 to 0.422) and 7.6\% (0.51 to 0.549) in terms of MAP and MRR when applied to 51 projects. 
	
	
	Note that the the idea of using an offline globally trained model for relevancy function can be added to all these five tools as well and potentially provide more total improvements over $BugLocator$.
	
	\subsection{Discussion from the perspective of the second heuristic: Word Embedding Technique}
	
	\begin{figure}[!t]
		\includegraphics[width=\linewidth]{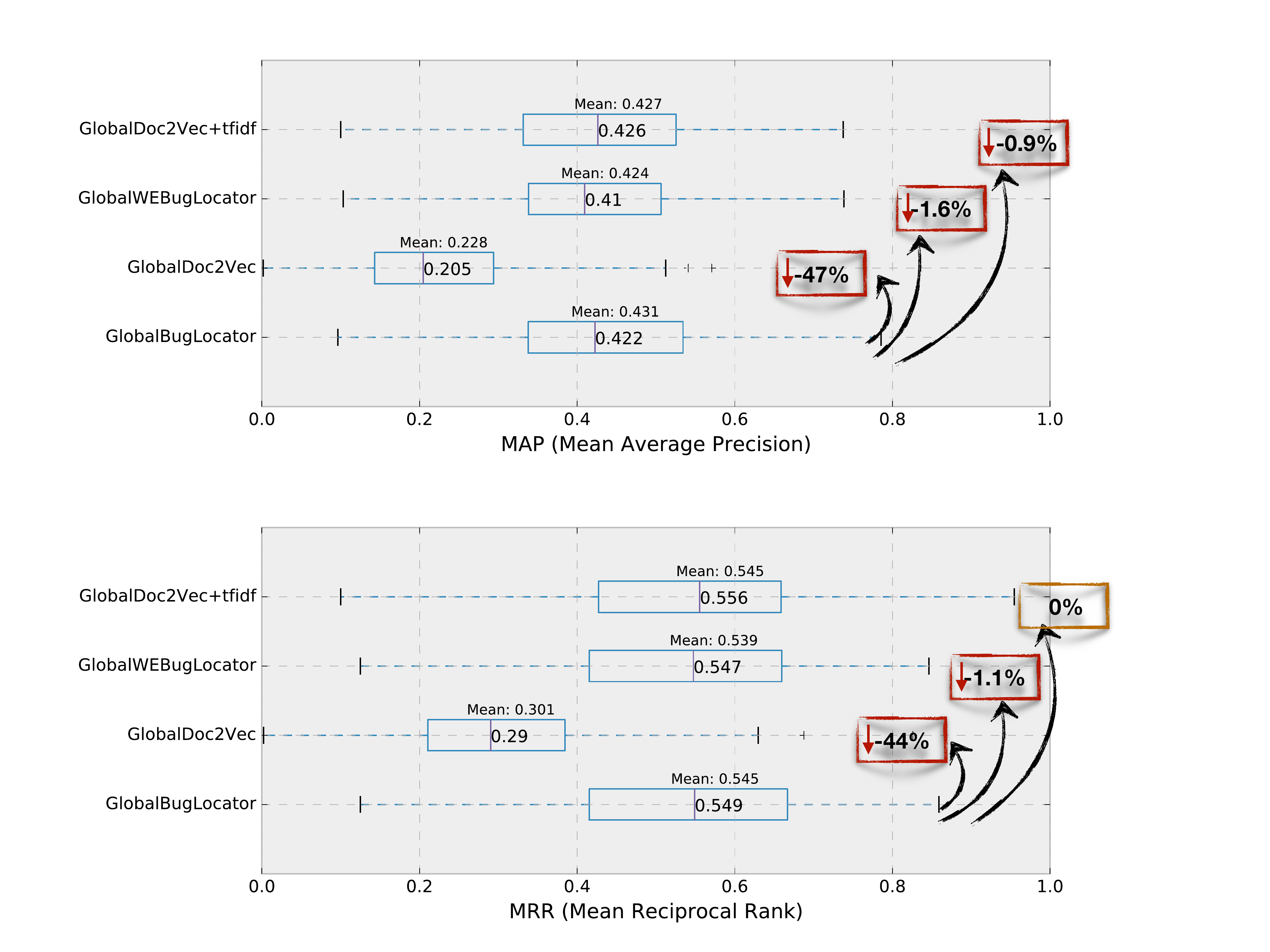}
		\caption{Performance of our second heuristic methods on the 51 projects}
		\label{Our Second Heuristic}
	\end{figure}
	
	
	Figure \ref{Our Second Heuristic} summarizes the performance result of the next three studied methods (Method 5, 6, and 7) and compare them with Method 4 ($GloBug$-Variation 1) in terms of MRR and MAP. As we see, Method 5 (Global Doc2Vec) with the lowest mean values of MRR and MAP, has the poorest performance among all four methods. Performance of Method 6 ($GloBug$-Variation 2) is very similar to Method 4 ($GloBug$-Variation 1). However, the mean performance of Method 4 is slightly better with rates of 1.6\% and 1.1\%, in terms of MAP and MRR. 
	
	Comparing performance results of Method 7 (the ``Combined'' method) and Method 4 ($GloBug$-Variation 1), we see that in terms of MAP Method 4 has a better mean performance with a rate of 0.9\% in the studied 51 projects. However, the median of Method 7 is 0.7\% higher. Also, in terms of MRR, we see that the two methods have the same mean performance. However, the median is 1.6\% higher in Method 7. 
	
	Therefore, the conclusion is that a complex Word Embedding method can potentially enable better exploitation of global data and as a result, outperform the global $TF.IDF$, however, this requires a systematic characterization of IRFL methods per projects, to understand in which projects such an embedding will be useful. In the following section, we take a look at one example scenario where Method 6 ($GloBug$-Variation 2) fails to improve the Method 4 ($GloBug$-Variation 1), and leave the systematic characterization study to future work.

	\subsubsection{An example scenario where simple is better than complex}

	\begin{figure}[!t]
		\includegraphics[width=\linewidth]{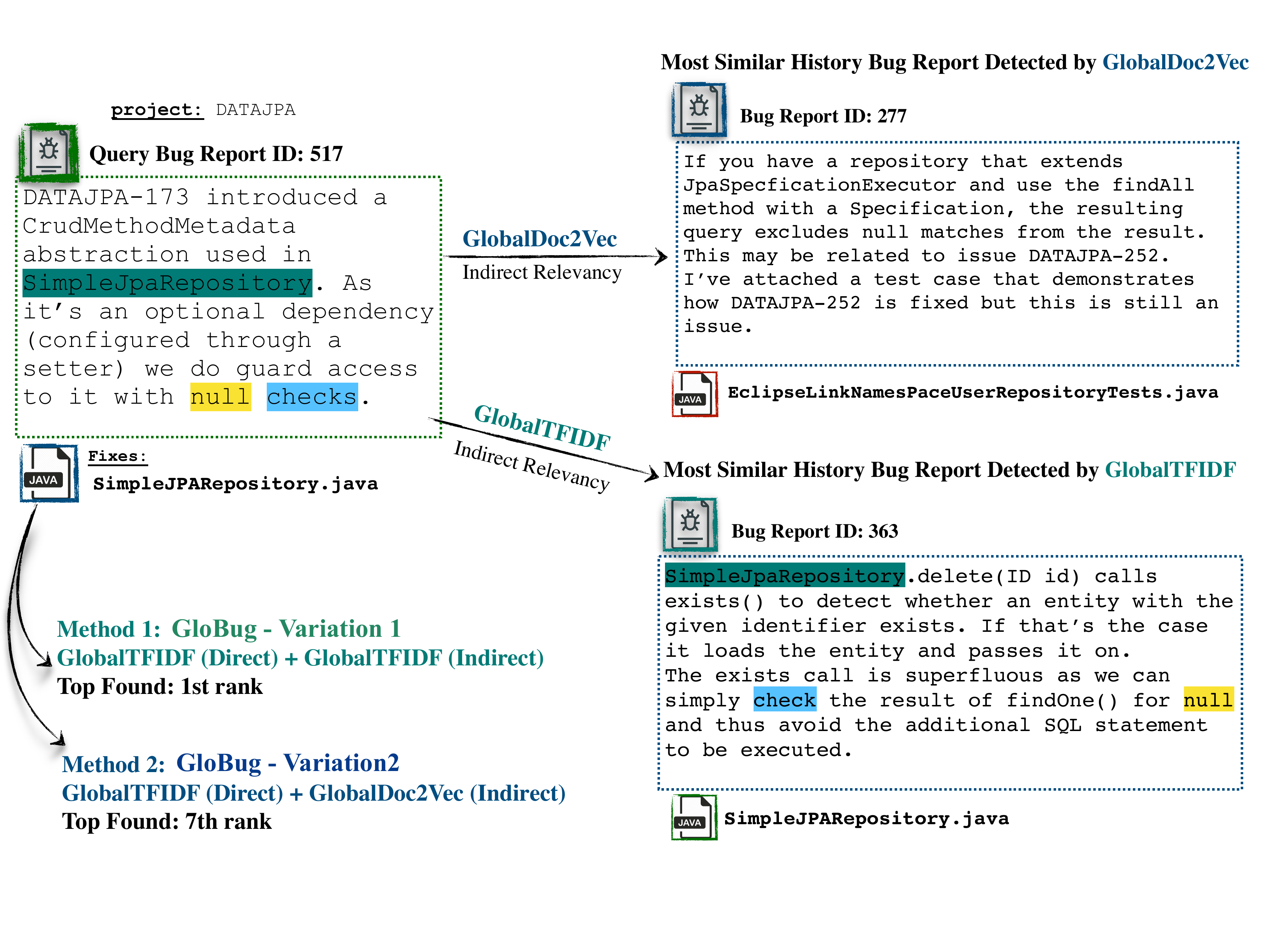}
		\caption{An example scenario where ($GloBug$-Variation 1) outperforms ($GloBug$-Variation 2)}
		\label{FailureScenario}
	\end{figure}
	
	Figure \ref{FailureScenario} represents an FL process where $GloBug$-Variation 1 (using global TF.IDF) outperforms $GloBug$-Variation 2 (using global Doc2Vec). This is a scenario in a software project called DATAJPA with 114 bug reports and a code repository of 330 Java files. The root cause of the bug is lying in a Java file called ``SimpleJPARepository.java''. Method 4 ($GloBug$-Variation 1) successfully locates the relevant file as its first ranked file, while Method 6 ($GloBug$-Variation 2) ranks the relevant file at its 7th rank.
	
	In this scenario, the global Doc2Vec model not only fails to improve the performance of the global $TF.IDF$, but also with adding too much complexity, it damages the performance of $TF.IDF$-based relevancy functions. This means that the semantics of the bug report, which is learned by the global Doc2Vec model is misleading the FL process. 
	
	\begin{figure}[!t]
		\includegraphics[width=\linewidth]{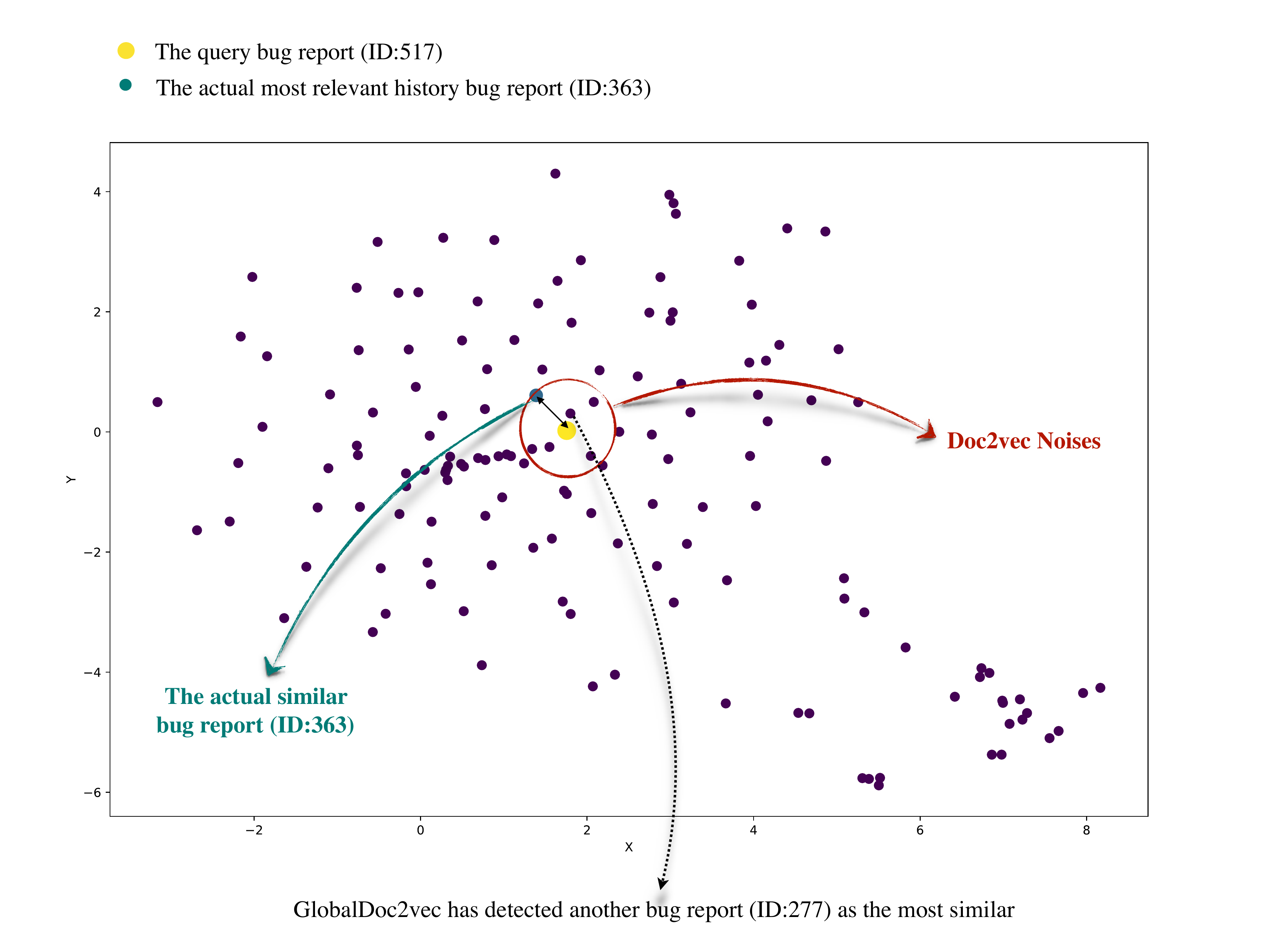}
		\caption{2-D Visualization (using TSNE) of Doc2Vec vector space for DATAJPA project with 144 bug reports}
		\label{2DExample}
	\end{figure}
	
	As mentioned, in $GloBug$-Variation 1, an indirect similarity score is calculated based on the common terms that appear in the two bug reports. As we see in this example, since the query bug report (ID:517) and its most similar history bug report (ID:363) both describe the problem with common program-related terms (highlighted), the Global TF.IDF is able to capture the correct relevance. However, in GLobal Doc2Vec the scoring process is much more complex than just looking at the common terms. In Global Doc2Vec, a neural network is trained using the co-occurrence of terms in all bug reports and source files of all global projects. Although this complexity enables a higher level of semantic comparison, in some cases such as this example, with too much noise added, it becomes misleading. 
	
	Although we cannot inspect why Doc2Vec is working this way in this example, due to the lack of explainability of neural networks, we visualize the Doc2Vec vector space of all 144 bug reports in DATAJPA in order to better understand the way that bug reports are ranked in this particular example. First, we reduced the Doc2Vec vector space into 2 dimensional space to be able to visualize the bug reports in 2-D space. Figure \ref{2DExample} depicts the 2-D visualization of Global Doc2Vec vector space for 144 bug reports of DATAJPA that was reduced from their original vector space into 2-D vector space using t-SNE \cite{tsne}. In this figure, the yellow dot denotes the query bug report and its actual most relevant history bug report is represented as a green dot. In this example, Global Doc2Vec is detecting 5 other bug reports more relevant to the query bug report and it has detected another bug report (ID:277) as the most relevant history bug report. The text of this bug report is represented in Figure \ref{FailureScenario}. 
	
	The reason why global Doc2Vec is able to contribute to the performance of the IRFL in some cases while it fails in some other cases, can be inspected in several factors. One of the most important factors is the quality of the bug report. In an IRFL process, bug report plays the most essential role. The effectiveness of the FL process is highly dependant on the quality of information (e.g. used language, keywords, stack-trace, etc.) that is provided in the bug report. For example, in this scenario, the person who has reported the problem seems to have very good knowledge of the software's internal structure. Therefore, he/she is describing the problem very precisely with terms that are specific to this software project.
	
	Based on the quality of a bug report, a certain IRFL method may be more suitable among other methods. For example, in this scenario, TF.IDF based solution will find several vocabulary matching between the bug report and the java file and as a result, ranks the relevant file at a very high position. But, Doc2Vec with adding too much complexity, disturbs the simple matching mechanism of the TF.IDF-based solution. 
	
	However, there are other factors involved in the determination of an IRFL performance as well. Size of a software project, size of the bug report repository, quality of other bug reports, pre-processing, and many other factors may influence the performance of an IRFL in a software project. 
	
	To summarize our second heuristic's discussion, we can say that a complex method is not always contributing to the IRFL performance. As Fu et. al. \cite{TimMenzies} also argue in their paper (``Easy over hard: a case study on deep learning''), sometimes simple algorithms work better than very complex models such as DNNs and complex algorithms should be applied cautiously in SE problems. 
	
	\section{Threats to validity}
	\label{threats}
	In terms of construct validity, we have used existing benchmarks, tool sets, and libraries to avoid implementation biases. We have also contacted authors of Bench4BL and confirmed our replication of their code-base and results. 
	
	Regarding internal validity, we have carefully designed multiple controlled experiments to separate the effects of relevancy function and the data-set, to avoid confounding factors when drawing conclusions.  
	
	With respect to conclusion validity, in each comparison, we run Wilcoxon Signed Rank test, which is a non-parametric paired hypothesis test, and report the P-values to make sure the conclusions are statistically valid. 
	
	Finally, in terms of external validity, we have used one of the biggest benchmark data-sets in fault localization studies to be able to draw more generalizeable conclusions. However, as always, it would be nice to be able to apply the findings on an industrial case study to see if the findings are only valid on open source systems or can be generalized to industrial cases as well. We also made sure to not generalize the results to any technique outside of what we have experimented with (Doc2Vec and TF.IDF).

	\color{black}
	\section{Related works}
	\label{RW_section}
	If we characterize $BugLocator$ as an IRFL technique that uses local TF-IDF and our study as a global TF.IDF plus a global embedding-based IRFL, we can classify the related IRFL work as those that use other relevancy functions than local TF.IDF and those that expand on the local TF.IDF implementation of $BugLocator$. 
	
	In the first category (non-TF.IDF-based relevancy functions), topic modeling is among most common ones. For example, Lukins et. al. \cite{LDA1,LDA2} presented an IRFL method based on Latent Dirichlet Allocation (LDA), where the vectorization is done using the LDA topic memberships per words. Although the application of topic modeling in FL was a novel idea at the time, it could never beat TF.IDF.
	
	In terms of using a neural embedding as the vectorizer, to the best of our knowledge, there is only one study by Xiao et al. \cite{deepLoc} in which they introduce DeepLoc. DeepLoc is composed of an enhanced CNN that considers bug-fixing recency and frequency, together with Word Embedding and feature detecting techniques. However, the results of their baseline ($BugLocator$) reported in this paper are not consistent with the original $BugLocator$ paper, which is what we have reproduced. Therefore, given that this paper's results were not reproducible, we could not compare our approach with DeepLoc.
	
	In terms of using Doc2Vec in other applications within software engineering, Doc2Vec has been successfully applied in the security domain for malware detection. For example, Ndichu \cite{d2v-JS} and Mimura \cite{d2v-VBA} have applied Doc2Vec to detect malware in Java Script and Visual Basic for Applications (VBA), respectively. Both studies show that regarding accuracy Doc2vec produced the best performance compared to other conventional language models in their experiment. 
	
	We also could not find any study that proposes a global corpus for the training phase in IRFL literature. 
	
	In the second category (those that extend $BugLocator$'s idea of a TF.IDF on bug reports and source code), however, there are many recent studies. All these techniques suggest extracting some extra data to leverage during the FL process.  For example, in 2013, Ripon K. Saha et. al. proposed BLUiR \cite{BLUiR}, which is built on top of $BugLocator$. BLUiR extracts code entities (e.g. classes, methods, and variable names) from bug reports and leverages them in the FL process. The authors showed that BLUiR outperforms $BugLocator$ in a set of 4 software projects. 
	
	In 2014, Chu-Pan Wong et. al. introduced BRTracer \cite{BRTracer}, which is also an extension of $BugLocator$. In some cases, stack traces are included in bug reports. So BRTracer analyzes stack traces shown in bug reports to improve bug localization accuracy.
	In the same year, Shaowei Wang et. al. introduced AmaLgam \cite{Amalgam}, which utilizes revision history in addition to similar reports and code entities.
	
	In 2015, Klaus Changsun Youm et. al. proposed BLIA \cite{BLIA}, which combines information such as similar reports, revision history, code entities, and stack trace information all together to improve the performance of $BugLocator$.
	
	In 2016, Ming Wen et. al. introduced Locus \cite{Locus}, the most recent technique, that leverages code change information to localize the buggy parts of the software.

	In another work, Xin Ye et. al. \cite{Learn2Rank} introduce an adaptive ranking approach that leverages domain knowledge through the functional decomposition of source code files into methods, API descriptions of library components used in the code, the bug-fixing history, and the code change history. Given a bug report, the ranking score of each source file is computed as a weighted combination of an array of features encoding domain knowledge, where the weights are trained automatically on previously solved bug reports using a learning-to-rank technique.

	In 2018, M. M. Rahman introduced BLIZZARD \cite{rahman2018improving}, which is an IR-based bug localization technique. In this study they locate buggy entities using appropriate query and an effective information retrieval technique. They use different reformulations for each type of bug report.

	In another work \cite{mills2018bug}, C Mills et. al. studied bug reports, to see if they have enough information for bug localization. In this work they studied the biases that may occur in the evaluation of IR-based techniques.  They claim that the bug report vocabulary is enough to formulate a query for bug localization.

	Recently in 2020, C Mills et. al. proposed an approach \cite{mills2020relationship} that uses genetic algorithm to extract a near optimal query from bug reports. Their GA select a subset of words from bug report to get better performance in IR-based methods, even when there is no localization hint in the bug report.

	Also, there are other works like \cite{ngram2012}, which have used n-gram method to improve information retrial. We can also mention \cite{spect-IR2015}, which has combined Information retrieval and spectrum based bug localization to get better results. 
	In addition, there are a few recent studies that use machine learning to train a supervised model for IRFL. Ngoc Lam et. al. present HyLoc \cite{IR-DL2017}, a direct-IRFL method that uses Deep Neural Network (DNN) in combination with rVSM. rVSM collects the feature on the textual similarity between bug reports and source files. Then, DNN is used to learn to relate the terms in bug reports to potentially different code tokens and terms in source files and documentation if they appear frequently enough in the pairs of reports and buggy files.
	
	The main difference between our proposed approach ($Globug$) and these related work is that $Globug$ can be considered as a complementary approach that can work along side many other heuristics. It also does not require extracting or preparing any extra information statically or dynamically. The only extra information $Globug$ uses is the freely available benchmark dataset (no preparation is required), which is used to train our language models only once, offline. The trained models (whether it is an embedding or a simple IDF) will then be reused to localize any new bug report on any project under study, with no extra overhead compared to basic $BugLocator$.  In other words, unlike some related work that require information such as stack trace data, revision history, or code change information, $Globug$ is as light-weight as $BugLocator$ and thus more applicable in industry were all the other extra requirements may not be always available. 
	
	\section{Conclusion and Future Work}
	\label{conclusion_section}
	
	This study was formed around two heuristics. First, we investigate the effect of a global training corpus on the performance of a state-of-the-art IRFL method in two modes of direct and combined (direct + indirect). Next, we study how a complex Word Embedding technique can be incorporated into an IRFL solution. We introduced $Globug$ with two variations to implement these heuristics and directly compared it with $BugLocator$ as a common baseline in IRFL. 
	
	The results showed that global data (heuristic 1) improves $BugLocator$ with an average rate of 14\% in terms of MRR and MAP in 64\% and 54\% of the cases respectively. 
	Also, heuristic 1 improved the mean performance of $BugLocator$ with average rates of 6.6\% and 4.8\% in terms of MRR and MAP when applied, on 51 software projects. We also showed that this amount of improvement is significant compared to the improvements of five other more recent IRFL methods provide over $BugLocator$. In addition, we discussed that our method does not require collecting extra data (such as change history or stack traces) per new project or new bug and works using a pre-trained model (one-time offline training) on a benchmark for all new projects and bugs. 
	We also showed that a complex Word Embedding solution such as Doc2Vec is not always effective and in some cases not only does not improve the performance of the IRFL method but also with adding too much noise, disturbs the performance of the simpler TF.IDF method.
	
	As a future study, we are planning to characterize the behaviour of the IRFL techniques, in the 51 software projects, to observe the factors that determine the performance of the FL methods. 
	The characterization takes place before applying the IRFL methods and is based on the current associations between source code files and bug reports, per project. It also leverages well-know static software metrics to characterize a project. Then predictive models are built based on this dataset to learn which type of projects require a more advance IRFL method and which ones are just fine with $BugLocator$.
	As explained in section \ref{discussion_section}, leveraging global data shows promising results in $BugLocator$ both in direct and indirect relevancy functions. Therefore, as another future work, we are planning to investigate global data on more existing IRFL techniques, specially those that are built on top of $BugLocator$. 
	Finally, we plan to use more recent word embedding techniques from NLP to see if the poor performance of Doc2Vec can be overcome for example with a context-ware model such as BERT\cite{bert}.
	


	%
	%

	

	\bibliography{mybibfile.bib}
	
\end{document}